\documentclass[screen,acmsmall]{acmart}

\copyrightyear{2021}
\acmYear{2021}
\setcopyright{acmlicensed}
\acmBooktitle{ACM Transactions on Software Engineering and Methodology}

\usepackage[english]{babel}
\usepackage{xcolor}
\usepackage{url}
\usepackage{listings}
\usepackage{tikz}
\usepackage{tikz-qtree}
\usetikzlibrary{trees}
\usepackage{graphicx}
\usepackage{xspace}
\usepackage{amsmath}
\usepackage{algorithm}
\usepackage[noend]{algpseudocode}
\usepackage{booktabs}
\usepackage{fancyvrb}
\usepackage{cleveref}
\usepackage{natbib}

\def\ninvalidvulnerabilities{69\xspace} 
\def\nvulnerabilitiesbeforefiltering{1,317\xspace} 
\def\nvulnerabilitiesindata{1,248\xspace} 
\def\nknownfixcommitsindatastart{2,080\xspace} 
\def\nunknownfixesfoundID{295\xspace} 
\def\nunknownfixesfoundNVD{16\xspace} 
\def\nknownfixcommitsindata{2,391\xspace} 
\def\precisionjimenez{22.36\xspace} 
\def\vulnerabilitiesfoundjimenez{279\xspace} 

\def\selectionrecall{93.59\xspace} 
\def\filterpercentage{11.00\xspace} 
\def\rankingrecallten{84.03\xspace} 
\def\rankingprecision{65.06\xspace} 

\def\realrankingrecallten{78.64\xspace} 
\def\realrankingprecision{60.89\xspace} 

\def\nevalcomponents{23\xspace}

\def\nmlmodelsevaluated{five\xspace}
\def\totalnumberofcommits{25,019,400\xspace} 
\def\candidatesinselection{2,753,058\xspace} 

\def\candidatesnotrelevant{422,312\xspace}
\def\percentageirrelevantextension{15.34\xspace}
\def\totalfilteringpercentage{9.32\xspace}
\def\nvulnerabilitiesinselection{1,166\xspace}

\newcommand{\toolname}{\textsc{FixFinder}\xspace}
\newcommand{\prospector}{\toolname}

\begin{document}

\begin{titlepage}


\enlargethispage{2\baselineskip}
\begin{center}
{\LARGE
Automated Mapping of Vulnerability Advisories \\
onto their Fix Commits in Open Source Repositories}\\[4mm]
{\large\bf [PRE-PRINT]} \\[6mm]
Daan Hommersom, Antonino Sabetta, \\
Bonaventura Coppola, Dario Di Nucci, Damian A. Tamburri
\end{center}

\thispagestyle{empty}
\vspace{8mm}
\small
The lack of comprehensive sources of
accurate vulnerability data represents a critical obstacle to studying and
understanding software vulnerabilities (and their corrections).
In this paper, we present an approach that combines heuristics stemming from
practical experience and machine-learning (ML)---specifically, natural
language processing (NLP)---to address this problem. Our
method consists of three phases. First, an \emph{advisory record} containing
key information about a vulnerability is extracted from an advisory
(expressed in natural language). Second, using heuristics, a subset of
candidate fix commits is obtained from the source code repository of the
affected project by filtering out commits that are known to be irrelevant
for the task at hand. Finally, for each such candidate commit, our method
builds a numerical feature vector reflecting the  characteristics of the
commit that are relevant to predicting its match with the advisory at hand.
The feature vectors are then exploited for building a final ranked list of
candidate fixing commits. The score attributed by the ML model to each
feature is kept visible to the users, allowing them to interpret the
predictions.

We evaluated our approach using a prototype implementation named
\prospector on a manually curated data set that comprises
\nknownfixcommitsindata known fix commits corresponding to
\nvulnerabilitiesindata public vulnerability advisories. When considering
the top-10 commits in the ranked results, our implementation could
successfully identify at least one fix commit for up to
\rankingrecallten{}\% of the vulnerabilities (with a fix commit on the first
position for \rankingprecision{}\% of the vulnerabilities). In conclusion,
our method reduces considerably the effort needed to search OSS repositories
for the commits that fix known vulnerabilities.  

\vspace{5mm}
\normalsize
\hrule
\vspace{2mm}
\begin{center}
{\large Citing this paper}
\vspace{3mm}
\end{center}


\noindent
Please cite this work as:
\vspace{3mm}
{\tt
\begin{Verbatim}
@misc{hommersom2021mapping,
  title  = {Automated Mapping of Vulnerability Advisories onto
             their Fix Commits in Open Source Repositories},
  author = {Hommersom, Daan and
            Sabetta, Antonino and
            Coppola, Bonaventura and
	    Dario Di Nucci and
            Tamburri, Damian A. },
  year   = {2021},
  month  = {March},
}
\end{Verbatim}
}

\vspace{3mm}
\hrule

\vspace*{\stretch{1}}
\includegraphics[width=4cm]{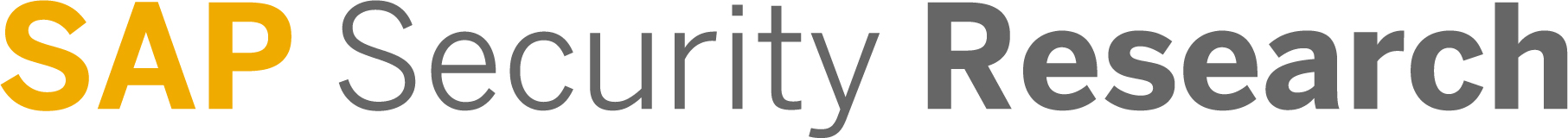}\hfill
\raisebox{-6mm}{\includegraphics[width=23mm]{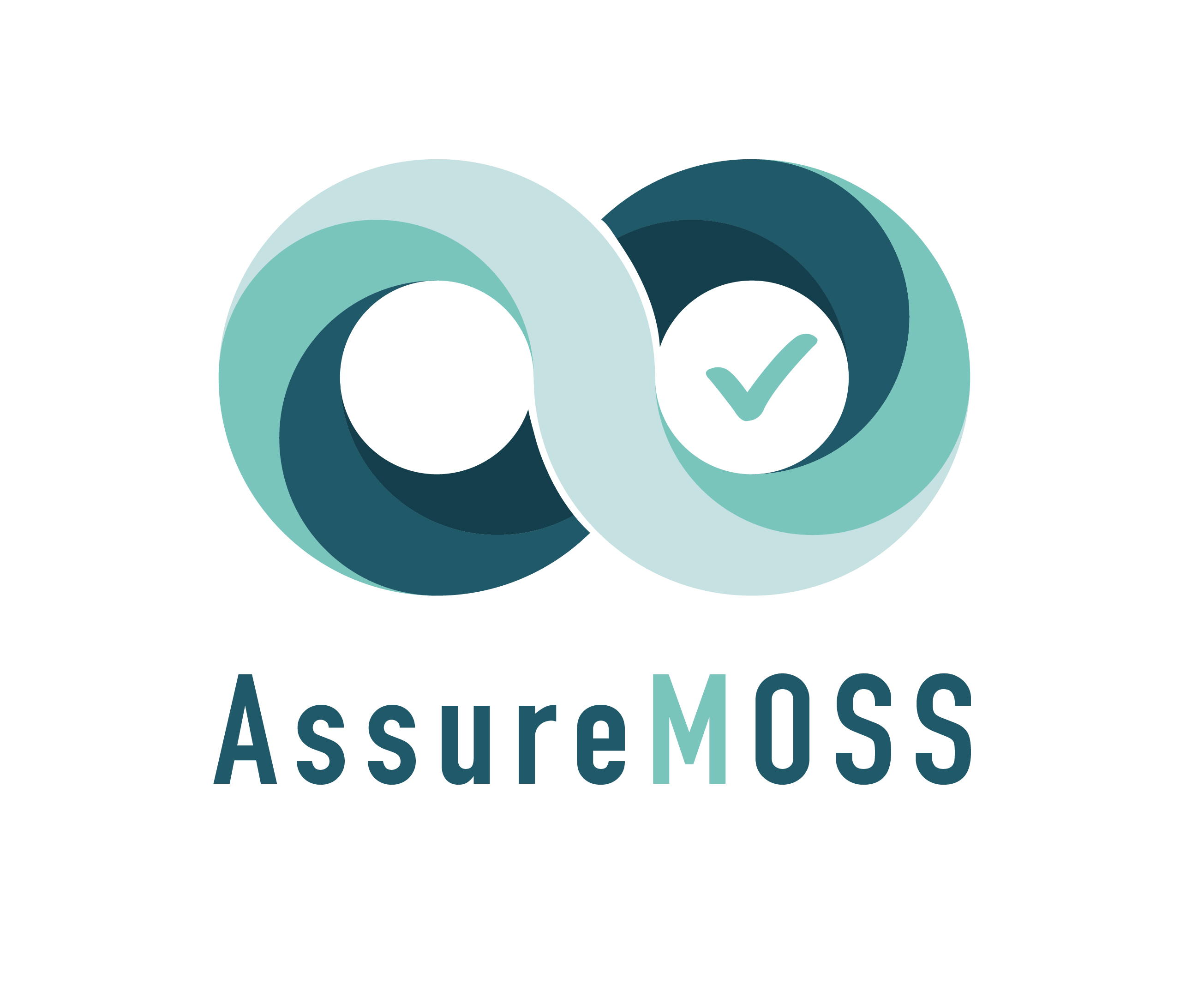}}


\end{titlepage}

\title{Automated Mapping of Vulnerability Advisories onto their Fix Commits in Open Source Repositories}

\author{Daan Hommersom}
\authornote{Daan Hommersom contributed to this work during his internship at SAP Security Research}
\affiliation{%
    \institution{Eindhoven University of Technology - JADS}
    \city{'s-Hertogenbosch}
    \country{The Netherlands}}
\email{daanhommersom@hotmail.com}

\author{Antonino Sabetta}
\affiliation{%
    \institution{SAP Security Research}
    \city{Nice}
    \country{France}}
\email{antonino.sabetta@sap.com}

\author{Bonaventura Coppola}
\authornote{Bonaventura Coppola contributed to this work while affiliated as Senior Researcher with SAP Security Research}
\affiliation{%
    \institution{University of Trento}
    \city{Trento}
    \country{Italy}}
\email{research@coppolab.com}

\author{Dario Di Nucci}
\affiliation{%
    \institution{University of Salerno}
    \city{Fisciano}
    \country{Italy}}
\email{ddinucci@unisa.it}

\author{Damian A. Tamburri}
\authornote{Damian A. Tamburri contributed in this work as Daan Hommersom's Final Thesis supervisor.}
\affiliation{%
    \institution{Eindhoven University of Technology - JADS}
    \city{'s-Hertogenbosch}
    \country{The Netherlands}}
\email{d.a.tamburri@tue.nl}



\begin{abstract}

    The lack of comprehensive sources of accurate vulnerability data represents a
    critical obstacle to studying and understanding software vulnerabilities (and
    their corrections). In this paper, we present an approach that combines
    heuristics stemming from practical experience and machine-learning
    (ML)---specifically, natural language processing (NLP)---to address this
    problem. Our method consists of three phases. First, we construct an \emph{advisory record}
    object containing key information about a vulnerability that is extracted from an advisory,
    such those found in the National Vulnerability Database (NVD). These advisories are
    expressed in natural language. Second, using heuristics, a subset of candidate
    fix commits is obtained from the source code repository of the affected project,
    by filtering out commits that can be identified as unrelated to the vulnerability at hand.
    Finally, for each of the remaining candidate commits, our method builds a numerical feature
    vector reflecting the  characteristics of the commit that are relevant to
    predicting its match with the advisory at hand. Based on the values of these feature vectors,
    our method produces a ranked list of candidate fixing commits. The
    score attributed by the ML model to each feature is kept visible to the users,
    allowing them to easily interpret the predictions.

    We implemented our approach and we evaluated it on an open data
    set, built by manual curation, that comprises \nknownfixcommitsindata known fix commits corresponding
    to \nvulnerabilitiesindata public vulnerability advisories. When considering
    the top-10 commits in the ranked results, our implementation could
    successfully identify at least one fix commit for up to
    \rankingrecallten{}\% of the vulnerabilities (with a fix commit on the first
    position for \rankingprecision{}\% of the vulnerabilities). Our evaluation shows
    that our method can reduce considerably the manual effort needed to
    search OSS repositories for the commits that fix known vulnerabilities.

\end{abstract}

\maketitle






\section{Introduction}

Over the past two decades, the availability of open-source software (OSS)
components and their adoption have grown so much that the applications that do
not include at least one open-source component have become the exception rather than the
norm~\cite{ZaffarKZ11}. According to a report by
Snyk~\cite{tal_state_2019}, the large majority of applications on GitHub
Enterprise~\cite{github_state_2020} include some open-source components, which
account for as much as 50\% of the entire code-base of those applications.

Because OSS components have become an integral part of the large majority of
software products (either free or commercial), their vulnerabilities can have a
disruptive impact on businesses and on the society at large.


The effective management of the software supply chain has proven to be a
critical problem of today's software industry and has received considerable
attention both from academia, from practitioners, and from tool makers, particularly
the numerous players in the Software Composition Analysis space that have emerged
in the past ten years.

While the early approaches (such as OwaspDC) attacked the problem of detecting
vulnerable dependencies by relying on metadata to link vulnerability advisories
(such as those collected by the National Vulnerability Database, NVD) and the affected
software artifacts (i.e. packaged dependencies), the limitations
of such approach soon became evident.

More advanced approaches in the literature use
code-level information about the vulnerable code fragment and its fix; this
enables a more accurate detection of vulnerable dependencies, as well as a precise
impact assessment through reachability analysis. Thew first of such approaches that go \emph{beyond metadata},
was first proposed, to the best of our knowledge, by Ponta et al. in 2015~\cite{ponta_emse_2020}.
While the public information about how the commercial tools that appeared in the market
later on is lacking, they often advertise capabilities that explicitly or implicitly
refer to a code-oriented approach.

Unfortunately, the data about security vulnerabilities is scattered across
heterogeneous sources, often not machine-readable, and does not provide the
the necessary level of detail, especially when it comes to code-level details. At
the same time, manual approaches~\cite{ponta_manually-curated_2019} to find such
data cannot scale proportionally to the amount of open-source code developed by
the community and incorporated in commercial products.

The role of automated tools becomes increasingly
important~\cite{sabetta_practical_2018,CabreraLozoya2021}. Academic and
industrial researchers have mostly focused on the classification of
security-relevant commits, where a critical piece of data is the commit that
introduces or fixes \emph{some} vulnerability. However, this kind of research
requires a large amount of training data to be available, which leads to a
somewhat circular problem.

To address this problem head-on, we argue that the first step is an automated
method to reduce the effort needed and increase the success rate in finding fix
commits for known vulnerabilities in OSS. While a completely automated method
that would ensure full accuracy is infeasible. We believe that an approach that
dramatically reduces the time spent by human security experts to mine source
code repositories to find fix-commits is of high practical value. To this aim,
we propose an approach consisting of three sequential steps:
\textbf{(1)}~\emph{extract} - a record containing all relevant vulnerability
information is created from the heterogeneous vulnerability data;
\textbf{(2)}~\emph{filter} - this \emph{advisory record} is used select
candidate commits by filtering out those that, based on heuristics motivated by
observations on the available vulnerability data, are known to be irrelevant for
the task at hand; \textbf{(3)}~\emph{rank} - these candidate commits are ranked
based on their probability of being a fix commit \emph{for the advisory record}
given as input.

In this paper, we present a prototype (named \prospector) that implements
our method. \prospector provides a proof-of-concept that can serve as a baseline
for further experimentation and research as well as for the implementation of a
usable solution to be adopted in an industrial context. \prospector is released
as open-source software as part of SAP's \textsc{project
    ``KB''}\footnote{\url{https://github.com/sap/project-kb}}, whose goal is to
\emph{support the creation, management, and aggregation of a distributed,
    collaborative knowledge base of vulnerabilities that affect open-source
    software}.

To the best of our knowledge, no similar research has been done in the current
state of the literature in the security engineering research domain, most of the
existing related research has rather focused on the related but distinct problem
of predicting whether a commit introduces or fixes \emph{some} vulnerability.
The problem of determining whether a commit is the fix commit \emph{for a
    specific vulnerability} (as described in an advisory) has not received much
attention from the research community, despite being of practical relevance.

Our experimentation with \prospector shows that this method can identify
the fix commits of an advisory with a precision of \rankingprecision{}\% and a
recall in the top ten (\rankingrecallten{}\%). This performance makes the tool
suitable for practical use in that it allows the effort of the human expert to
be dedicated to the (relatively few) residual cases that the tool could not deal
with.

The remainder of the paper is organized as follows. In~\Cref{motivation}, we
provide some background information and elaborate on the need for a method to
automate the process of finding fix commits for known vulnerabilities in OSS.
\Cref{description} describes the three pillars of the approach; our
implementation of this method is presented in \Cref{implementation}. The
evaluation of our method is presented in \Cref{evaluation}. In
\Cref{discussion}, we elaborate on the findings gathered from our experiment,
and we outline some opportunities for further research. After a brief overview
of the current state of the literature is given in \Cref{related_work},
\Cref{conclusion} concludes the paper.

\section{Background and Motivation}
\label{background_and_motivation}

\subsection{Vulnerabilities of Open Source Software}
\label{introduction_vulnerabilities}

According to the European Union Agency for
    Cybersecurity~\cite{enisa_vulnerabilities_nodate}, a security vulnerability
    is \emph{a
    weakness an attacker could take advantage of to compromise the confidentiality,
    availability, or integrity of a resource}. Vulnerabilities can allow malicious
parties to steal, destroy or modify sensitive data, causing disruption in the
form of data loss, system downtime, etc.
Depending on the severity of the vulnerability, the level of damage can be
enormous. The damage done is twofold: it affects customers as their personal
data such as credit card numbers and passwords can be stolen, and it affects
businesses as their data can be stolen and their services can be disrupted.

The adoption of software reuse, particularly of third-party libraries released
under open-source licenses, has dramatically increased over the past two
decades. It has become so pervasive in today's software, including commercial
products, that the once clear-cut distinction between proprietary and
open-source software has gradually blurred. Whereas traditionally software
vendors used to have full control on the entire development process of most of
the components that made up their products, nowadays a large part of the
codebases of those products come from community-developed free open-source
(FOSS) projects, managed by independent parties, each with their own
unsynchronized lifecycles, heterogeneous quality standards, and development
practices. Building upon these free, high-quality, community-developed building
blocks, vendors can focus their efforts on differentiating features and deliver
innovative capabilities faster. When doing so, however, they become responsible
for assessing and mitigating the impact that a vulnerability in those
open-source components might have on their products.
On average, open-source projects have 180 package dependencies, and a large
number of projects can depend on one project~\cite{github_state_2020}: a
vulnerability in any one of these dependencies can enable malicious parties to
do harm. For example, in the \textit{npm} ecosystem, the average package has 3.5
million dependent projects but fewer than 40 direct contributors. Moreover,
37\% of open-source developers do not implement any sort of security testing
during development, while 81\% of the users feel developers are responsible for
open-source security, and only three in ten open-source maintainers consider
themselves to have high security knowledge~\cite{tal_state_2019}.


\subsection{On the Need for Code-level Data about Vulnerabilities and their Fixes}
\label{previous_method}

The existing standard for reporting vulnerabilities is the Common
Vulnerabilities and Exposures (CVE) identifier. CVEs are assigned by MITRE, a
research organization funded by the U.S. government. The largest database of CVE
vulnerability data is the National Vulnerability Database (NVD) from the U.S.
National Institute of Standards and Technology (NIST). The NVD standardizes the
CVE published by MITRE in a data format that enables researchers to directly
process their data, and provides an API to query this data.

To enable secure usage of OSS, product development teams are required to
frequently scan the  open-source components their project depends upon and
ensure that they are not affected by known vulnerabilities. Because mapping
software artifacts as used in development environments with the metadata in
advisories is difficult and potentially unreliable, tools such as
\textsc{Eclipse Steady}\footnote{\url{https://github.com/eclipse/steady}} use
code-level analysis to determine which software artifact is affected by which
vulnerability. To do so, Eclipse Steady relies on the availability of accurate
data about the code-level changes (commits) that fix each known vulnerability.
To support the tool operation, SAP has manually created a knowledge base of
vulnerabilities that affect open-source components used in SAP software (either
products or internal tools). Early 2019, this knowledge base was released on
GitHub as \emph{project
``KB''}~\cite{sap_sapproject-kb_2020,ponta_manually-curated_2019}.


Such commits are used to determine if any dependency of a subject project
contains the vulnerable code fragment, and if that code fragment can be actually
reached, via concrete executions or static reachability
analysis~\cite{ponta_emse_2020}.

While one of the advantages of OSS is free availability, accurate (code-level)
data about its vulnerabilities and fixes are hard to obtain. These data are
often scattered across different sources: public vulnerability databases such as
the NVD, project-specific issue trackers, websites publishing security
advisories and proprietary databases such as those offered by vendors of
software composition analysis (SCA) tools.

This difficulty in obtaining vulnerability data hinders further development of
new tools that could push the state of the art in vulnerability detection and
mitigation. Also, a considerable effort is spent by multiple parties to
independently search for vulnerability data, without addressing the problem at
the root.

When initiating a vulnerability analysis from the largest vulnerability database
(NVD), finding the commit that fixes the disclosed CVE may be a difficult task;
in fact, the fix commit may often not be found at all. This is emphasized
by~\cite{schryen_is_2011}, which reports that for 17.6\% of the published OSS
vulnerabilities no patch information can be found. To demonstrate that one
cannot solely rely on NVD advisories to find fix commits, we performed an
analysis of the coverage of the NVD for the fix commits in our dataset.

Jimenez et al.~\cite{jimenez_enabling_2018} have proposed an approach to
automatically find fix commits for CVEs published in the NVD. Their method
relies on two sources of information: (i) CVE IDs in commit messages and (ii)
links referred to in NVD. For the latter, their tool checks if the NVD page
contains a link to a commit in the repository. However, often NVD does not refer
to a fix commit at all, and in practice, there is no one-to-one correspondence
between a commit and a vulnerability.

\section{Solution Design: General Overview}\label{description}

\begin{figure}[ht]
    \centering
    \fbox{
        \includegraphics[width=0.8\textwidth]{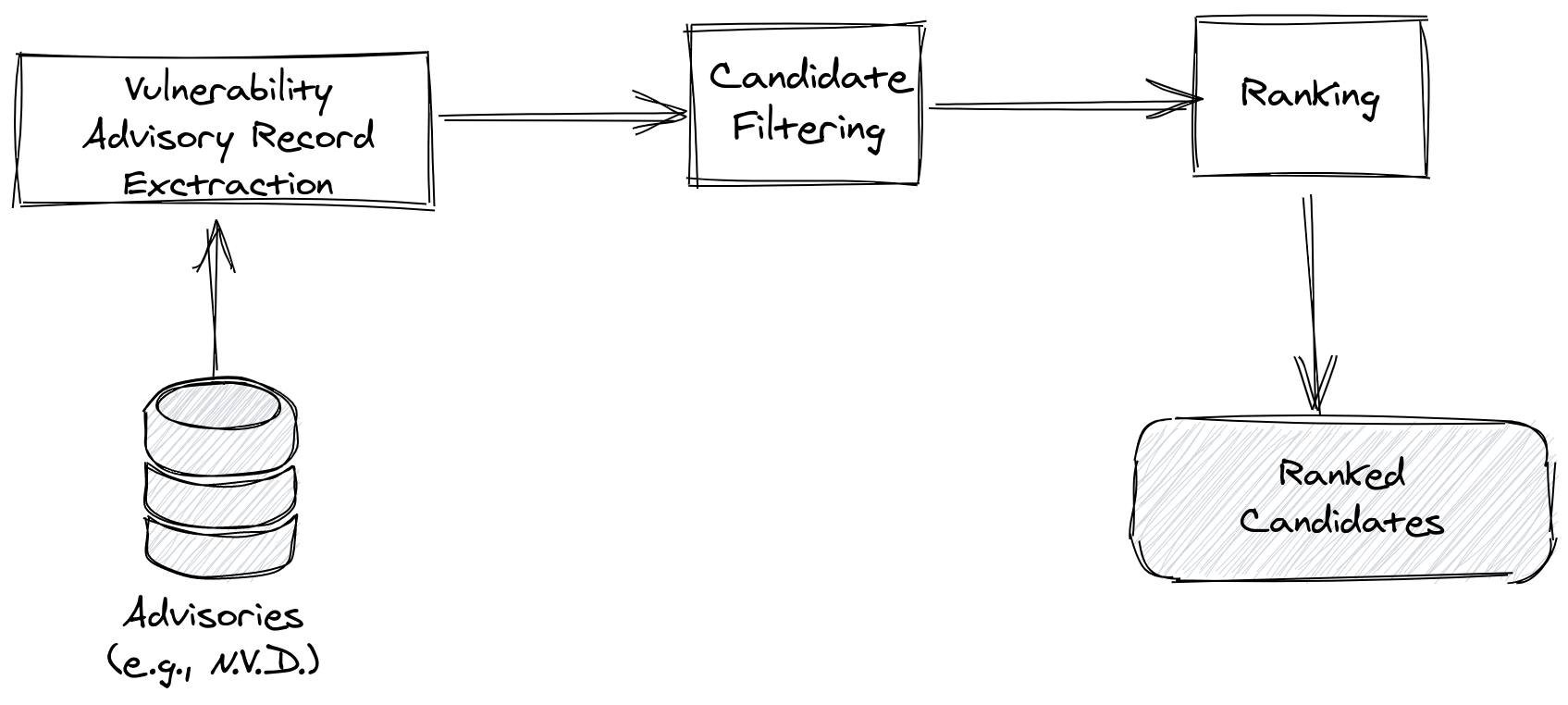}
    }
    \caption{High-level overview of the approach}
    \label{fig:overview}
\end{figure}

We propose an approach of automating the process of finding fix commits,
inspired by the steps that a security researcher would perform manually. Such an
approach is therefore based on three phases, described below, that are
functional to reducing the research problem introduced previously into a problem
which can be addressed via machine-learning. A high-level outline of the
three-phase approach is depicted in \Cref{fig:overview}.

\subsection{Phase 1: Information Extraction}\label{experimental_setup_extract}
The first issue within this phase is to ensure the extraction only the relevant
information from all available sources. To this purpose, in this first phase, we
create an \emph{advisory record}, containing all relevant information for a
vulnerability expressed in natural language, such as the vulnerability
description and the repository URL.

When searching for fix commits for a vulnerability which has been assigned a CVE
and is published in the NVD, the NVD can be used to extract data as the
vulnerability description, publication date and references.

\begin{figure}[ht]
    \centering
    \fbox{
        \includegraphics[width=0.95\textwidth]{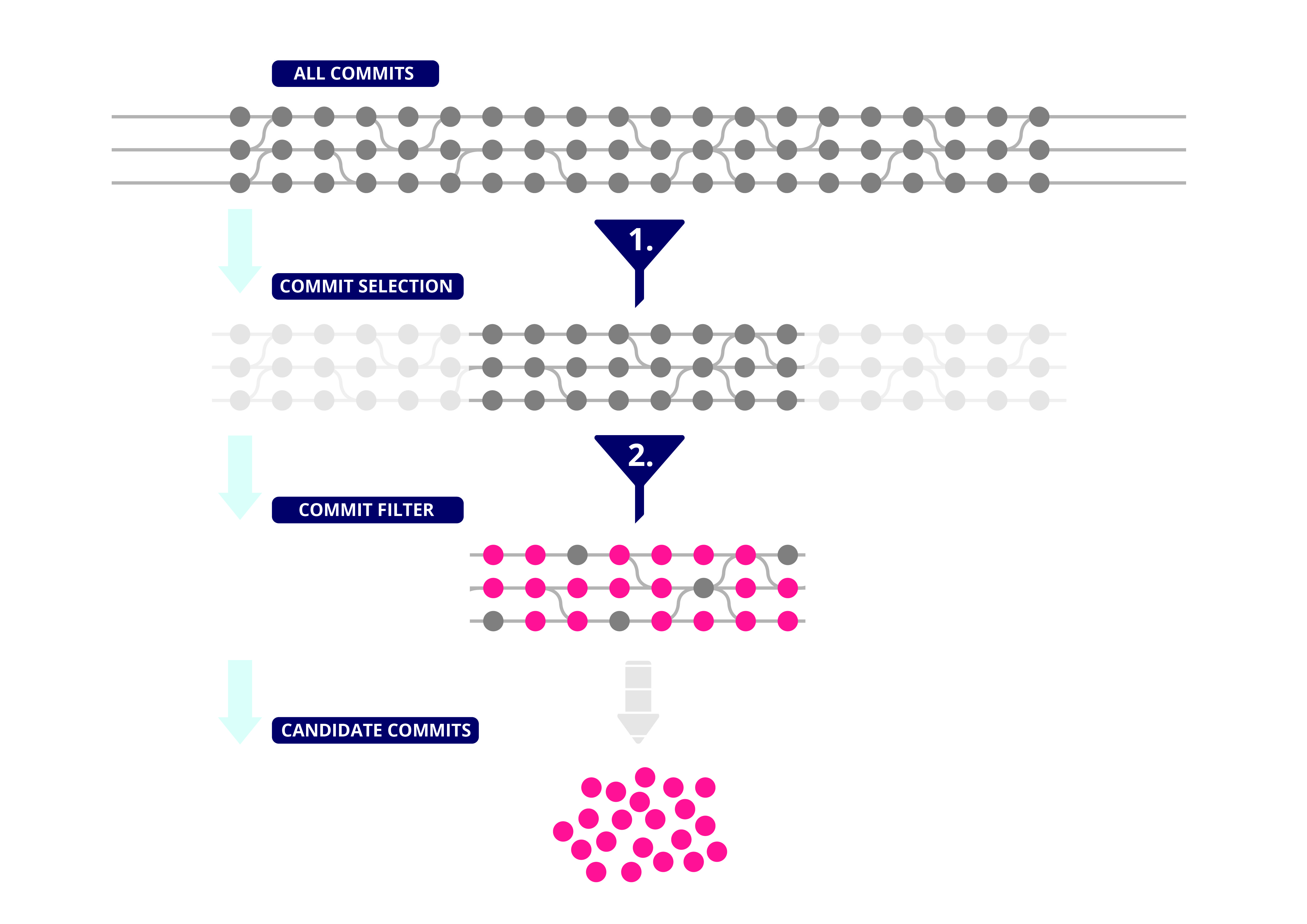}
    }
    \caption{The filtering procedure: The large number of commits is reduced
        to a set of candidate commits through selecting a subset and filtering
        irrelevant commits from this subset.}
    \label{fig:filtering_visualized}
\end{figure}

\subsection{Phase 2: Candidate Filtering}\label{experimental_setup_filter}
After all valuable information is extracted, such data can be used to identify
fix commits. One could regard every commit in a repository as a commit that
could be the fix commit. However, as repositories can have a large number of
commits (e.g., the git repository of the Linux
kernel\footnote{\url{https://github.com/torvalds/linux}} has almost 1M commits
at the time of writing, with more being added on a daily basis), identifying the
fix commit among this large number of commits would be impractical. Therefore,
the next step is to come to a selection of \emph{candidate commits}, obtained by
discarding those that are known to be irrelevant for the vulnerability advisory
at hand.

The filter phase has as the objective to reduce the number of commits to
consider as a candidate commit without discarding fix commits, which is a
trade-off between precision and recall. The larger the number of candidate
commits (low precision), the greater the odds that the fix commits are in
such selection (high recall). On the other hand, the smaller the number of
candidate commits (high precision), the higher the chance of discarding also
commits that happen to be precisely the fix commits one is looking for.
\Cref{fig:filtering_visualized} depicts this procedure.


\subsection{Phase 3: Candidate Ranking}\label{experimental_setup_rank}

As a result of the previous phase, a set of candidate fix commits is
determined. The third and last step now consists in ranking such candidates into
a list of commits sorted by decreasing probability of being true fix
commits. Finally, the most likely fix-commits will appear at the top
of the list. This is a standard ranking task which can be approached
exploiting a number of machine learning (ML) algorithms. As a preliminary step, a
feature space is defined, so that each candidate commit is represented as a vector of
informative features. Eventually, these feature vectors constitute the
input of a ranking algorithm that assigns a score to each candidate.
The task of selecting informative features makes the most relevant part of a
ML approach and is described in detail in
\Cref{implementation}. In order to train the learning algorithms, a
training set must be built including both fix commits as positive training
examples, and non-fix commits as negative training examples, drawn from the same
repository~\cite{sabetta_practical_2018} and respecting their naturally
occurring distributions.

\section{Prototype implementation: \prospector}\label{implementation}

\prospector\footnote{\prospector is available
    under the terms of the Apache 2.0 version from the GitHub repository of
    \textsc{project ``KB''} \url{https://github.com/sap/project-kb}}, is a prototype
implementation of the approach described in \Cref{description}. In the
following sections, we detail how we extract the vulnerability information from
an advisory, how we prune the large initial number of candidate commits, how we
obtain a suitable feature vector representation for them, and finally how we rank the
remaining candidate commits.

\subsection{Design of \prospector: a Data-Driven Approach} In order to
evaluate our approach, we use the publicly available vulnerability knowledge
base disclosed by Ponta et al.~\cite{ponta_manually-curated_2019}, extended with
additional records that were collected at SAP Security Research after the
publication of~\cite{ponta_manually-curated_2019}\footnote{These data are still
not published at the time of writing, but there are plans for SAP to add them to
project ``KB'' in 2021}. Our dataset consists of \nvulnerabilitiesindata
vulnerabilities with vulnerability identifiers (referred to as CVEs),
vulnerability descriptions, and a total of \nknownfixcommitsindata known fix
commit identifiers. While only part of this dataset is currently public, our
study can be partially replicated using the dataset by Ponta et
al~\cite{ponta_manually-curated_2019}, as we show in \Cref{dataset_discussion}.

From the initial set of \nvulnerabilitiesbeforefiltering vulnerabilities, we
removed \ninvalidvulnerabilities vulnerabilities for which the affected
repository could not be cloned or none of the fix commits mentioned in the
vulnerability statement could be obtained after cloning the entire
repository\footnote{This is due, for example, to commits in our training dataset
that belonged to branches that do not exist any longer, making the commit
unreachable (and subject to Git's garbage collection).}. This eventually led to
a dataset including \nvulnerabilitiesindata unique vulnerabilities.

There are two common procedures to automatically find fixes for vulnerabilities:
through following references in the NVD, or through finding a vulnerability ID
in the commit message~\cite{jimenez_enabling_2018}. There are
\nknownfixcommitsindatastart fix commits known for the \nvulnerabilitiesindata
vulnerabilities in the dataset. We added an additional \nunknownfixesfoundID
commits as the commit messages contained the vulnerability ID of one of the
vulnerabilities in the data, and another \nunknownfixesfoundNVD commits as the
NVD referred directly to these commits. Using this technique, we grew our data
set to \nvulnerabilitiesindata vulnerabilities with \nknownfixcommitsindata fix
commits.

Since a large number of commits need typically to be compared, we create a data
base to store the commit meta-data and content; the commit timestamp, the commit
message, the changed files, and the git diff content. The git diff shows what
changes have been made in the commit, and compares the two states of the files
that have been changed. The information about the two files which are compared
was already captured in the 'changed files' variable thus is not needed.
Therefore, a line was only kept if it did not start with \texttt{'diff --git'},
\texttt{'index '}, \texttt{'+++ '}, \texttt{'--- '} or \texttt{'@@ '}, whereby
noise was reduced.

For every column in the database, a pre-processed column is added. The content
is pre-processed through removing all tokens that do not contain at least one
alphabet letter (i.e., only digits), and tokens that are just one character. The
SpaCy NLP Library~\cite{spacy} is used to tokenize the content, remove stopwords
(such as determiners and prepositions) and splitting tokens that are either in
\emph{CamelCase}, \emph{snake\_case}, or \emph{dot.case}. All resulting words
are lemmatized using SpaCy. Finally, all tokens are converted to lowercase.

\subsection{Core Vulnerability Information Extraction}
The \prospector workflow begins with an information extraction stage
focused on selecting and extracting from advisory pages those pieces of
vulnerability information which are deemed both relevant and reliable towards
finding fix commits. The considered sources are the NVD and the advisory pages
pointed by the NVD. This process is actually an intermediate step towards
generating the features supporting the final ML-based ranking task. For every
CVE, the selected information pieces are collected into an advisory record that
is the distilled, structured version of the available information we could
reasonably find about such individual CVE. Since our dataset contains
vulnerabilities that have been published in the NVD, we use the new NVD CVE/CPE
API to extract data as the vulnerability description, publication date and
references. The NVD often refers to advisory pages that are relevant for the
given vulnerability. Therefore, we explore how these references can be utilized
when trying to find fix commits. To test this, we crawl all references and
extract their content, such as all URL references on the sites. We then use this
information to create features from. Later on, we will inspect the weights the
ML models assign to these features in order to check whether these features
bring a positive impact on the predictions.

\begin{figure}[ht]
    \centering
    \begin{minipage}[b]{0.85\textwidth}
        \includegraphics[width=\textwidth]{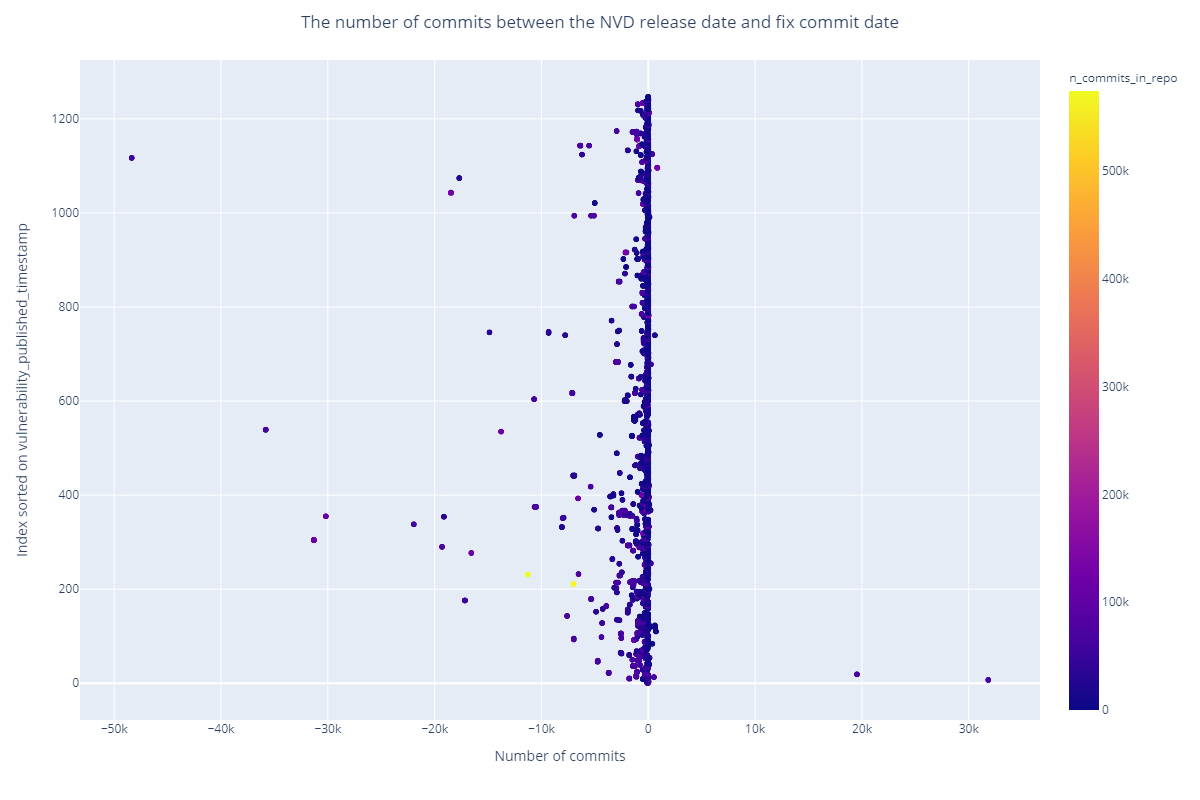}
    \end{minipage}
    \\
    \begin{minipage}[b]{0.85\textwidth}
        \includegraphics[width=\textwidth]{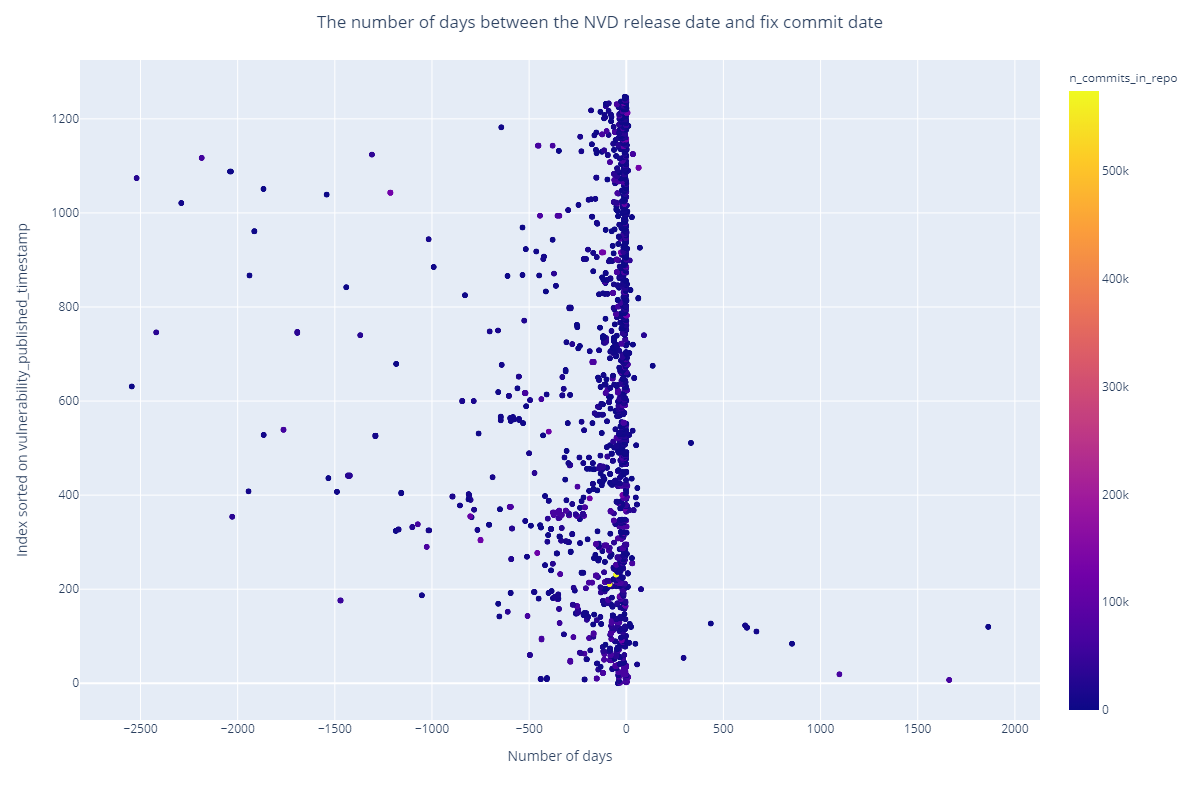}
    \end{minipage}
    \caption{The distance between the fix commit timestamp and the CVE publication
        timestamp expressed in the number of commits (top) and the number of days (bottom).}
    \label{fig:selection_commit_distance}
\end{figure}

\subsection{Filtering of fix-commit candidates}
In order to reduce the amount of processed data, we choose to perform a
reasonably safe preliminary pruning on the whole set of commits. We select a
subset of the commits in the repository based on the NVD publication date of the
vulnerability. In addition, we discard commits that do not change at least one
code file by means of filtering on extensions (e.g., commits that just change the
documentation of the project). The process eventually generates the actual set
of commits that we treat as candidate fix commits.

To determine the strategy for filtering the commits based on the distance from
the CVE publication, we computed the difference between the fix commits and the
CVE publication date and plotted it in \Cref{fig:selection_commit_distance}. The
commit distance is the number of commits that can be found between the fix
commit timestamp and the CVE publication timestamp, and the days distance is the
number of days between these two timestamps. When the number is negative, this
means that the fix commit was before the CVE publication. When the number is
positive, the fix commit was after the CVE publication. The y-axis reports the
indices of the vulnerabilities sorted by publication date of the vulnerability,
thus y = 0 stands for the oldest vulnerability in the data.

\begin{figure}
    \hspace*{-6mm}
    \includegraphics[width=1.07\textwidth]{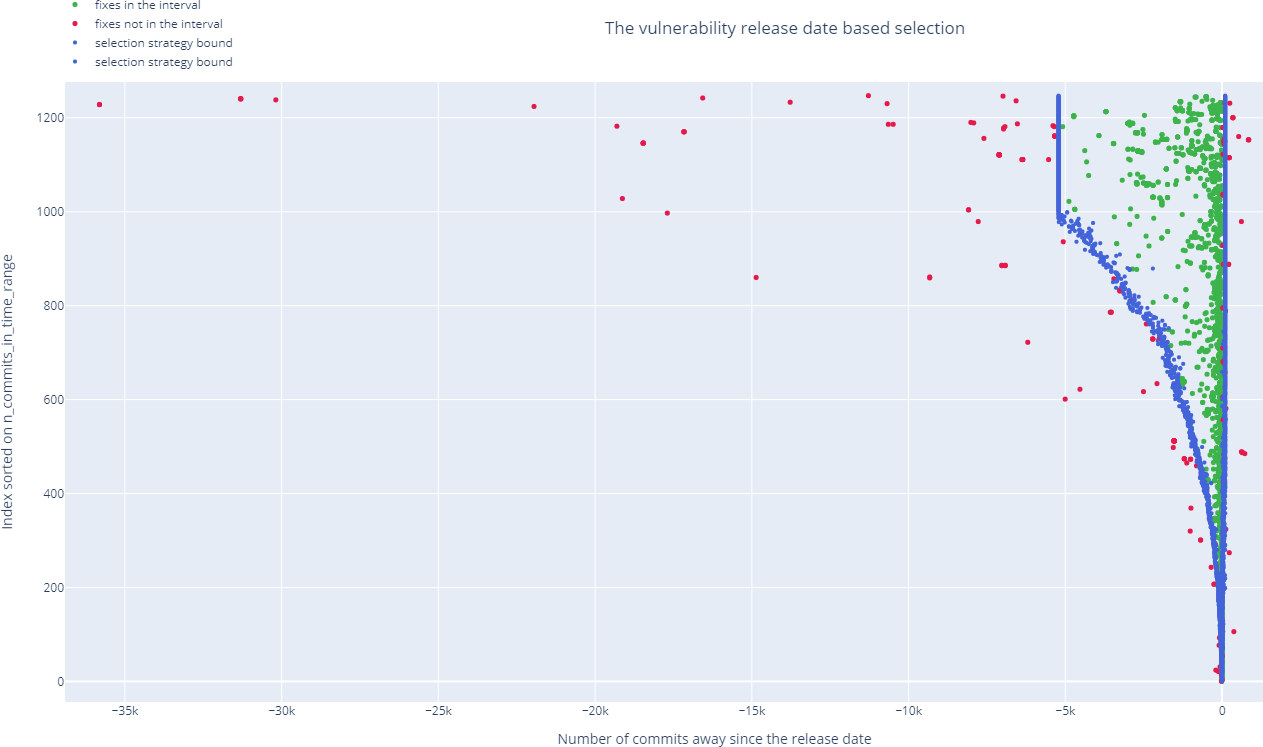}
    \caption{The effect of filtering commits based on their distance from the CVE publication, based on a combination of the number of days and the number of commits. The y-axis corresponds to the vulnerabilities sorted on the number of commits that are within two-years before the release date and one-hundred days after, so all dots with the same y-value correspond to the same vulnerability. The x-axis is the number of commits between the dot and the vulnerability release date. The red and green dots correspond to known fix commits, where the green dots are fix commits that fall within the selection. The number of commits that is in the selection, based on a combination of time and a maximum number of commits, is visualized by the blue dots. So for every blue dot, there is at least one green or red dot on the same y-value.}
    \label{fig:selection_visualized}
\end{figure}

From \Cref{fig:selection_commit_distance}, we can observe that various fixes are
far away from the CVE publication date. Therefore, trying to determine an
interval that captures all commits, based on the CVE publication date, would
result in a too large number of candidate commits, which would be impractical to
treat. Please consider that this filtering could hinder the recall of the
approach. Nevertheless, the interval should be selected to find the correct
balance between lower recall and higher scalability. To come to an initial
selection of candidate commits, we select all commits within two years before
and one hundred days after the release date with a maximum number of commits of
respectively 5215 and 100 commits. This selection strategy, visualized in
\Cref{fig:selection_visualized}, results in a selection recall of
\selectionrecall{}\%  but reduces the number of commits to
\filterpercentage{}\%.

After we determine this commit interval, we select all commits within it. From
this selection, commits that do not change the actual code can be discarded
since a fix commit should at least change one code file. This can be achieved
through the criterion that a commit should have changed at least one file with a
relevant extension.
The exact list of those considered in our implementation is: \texttt{["java",
            "c", "cpp", "h", "py", "js", "xml", "go", "rb", "php", "sh", "scale", "lua",
            "m", "pl", "ts", "swift", "sql", "groovy", "erl", "swf", "vue", "bat", "s",
            "ejs", "yaml", "yml", "jar"]}.

For every candidate commit, we check whether the commit changes at least one
file with a relevant extension through splitting the filenames on the dots (e.g.
\texttt{'example.py'} becomes \texttt{['example', 'py']}) and checking whether
the last element of one of these files is in the list of relevant extensions. If
the commit does not change a relevant file, we filter it from the selection. We
regard the remaining commits as candidate commits.

\subsection{Feature Engineering for the commit ranking task}
For every candidate commit, we create a corresponding feature vector, which
constitutes our set of \textit{independent variables}. This feature vector
comprises multiple component features that take numeric values providing
information on the candidate commit. In total, \nevalcomponents features are
extracted to predict the probability of a candidate commit being a fix commit
and sort the commits based on this probability score, which constitutes our
\textit{dependent variables}. For this prediction task, we evaluate
\nmlmodelsevaluated ML models. The features used as \textit{independent
variables} by the model and needed for the ranking task are the following.

\begin{description}
    \item[Vulnerability ID in message.]\label{vulnerability_id_in_message} As
          explained in \Cref{motivation}, one way to find fix commits for a
          given vulnerability is to search for commits that include the CVE
          vulnerability identifier in the commit message. Therefore, we create a
          feature that contains this information. When the vulnerability ID is
          found in the message the feature value equals 1.0, otherwise the value
          equals 0.0. It is important to emphasise that 
          the goal of our approach is not to find \emph{any} fix-commits (that is,
the fix commits of any vulnerability), but rather to find the particular commits that
fix \emph{the vulnerability at hand}\footnote{In other words, if our approach is used to find the commit(s) fixing vulnerabiilty
$V_1$ and it comes across candidate commit $C$ whose commit message says ``This is the
fix to security vulnerability $V_2$''. Reporting $C$ as a fix for $V_1$ would be wrong:
generally speaking, $C$ is definitely a security fix, but not the one we'd be looking for.}.
          For this reason, we add another feature that indicates whether
          a vulnerability ID, other than the one of interest provided as input, is
          mentioned in the commit message of the candidate. The value of this additional feature  is set
          to 1.0 when the identifier of the vulnerability of interest is not in the commit message but the
          string \texttt{'CVE-'} is present instead (indicating that some other vulnerability is mentioned),
          otherwise it is set to 0.0.

    \item[`Referred to' features.]\label{rvd_referred_to} The second method to
          find fix commits that is proposed by Jimenez et
          al.~\cite{jimenez_enabling_2018}, and adopted by others, is through
          following the hyperlink Web references in the NVD. When a NVD entry
          refers to a commit in the repository of the affected project, such
          commit is labelled as a fix commit. Therefore, we create a feature
          that reflects whether the references in the advisory record contain a
          reference to a commit. In practice, when a URL contains the string
          \texttt{'/commit/'}, the first eight characters after this pattern are
          stored in a list. Then, for all candidate commits, we check whether
          the first eight characters of the commit ID can be found in this list.
          If this is the case, the value for these feature equals 1.0, and 0.0
          otherwise. As it can be seen in \Cref{tab:nvd_reference_analysis} in
          \Cref{motivation}, only 7.53\% of the known fix commits is referenced
          by the NVD. Therefore, we create another feature which reflects
          whether the commit is referred to by an advisory reference. By means
          of web-scraping, we extract all URLs from the references in the
          advisory record (advisory pages). We then recursively apply the same
          approach as described above to provide a score that reflects whether a
          commit is mentioned  by one of the advisory references.

    \item[Commit statistics.]\label{rvc_commit_statistics} There are some commit
          statistics that can be used as a predictor for a commit being a fix
          commit, such as the spread of the patch~\cite{sawadogo_learning_2020}.
          A commit can change an arbitrary number of files and an arbitrary
          number of lines which can be represented as the spread of the patch:
          the more files and continuous blocks of lines (hunks) have been
          modified, the more spread out the patch is. Therefore, we create three
          different individual features indicating the spread of the patch: the
          number of hunks, the average hunk size, and the number of changed
          files.

    \item[Commit message references.]\label{rvc_message_references} On GitHub,
          developers can create `issues' to report bugs such as vulnerabilities
          or `pull requests' to propose modifications to a repository. When a
          commit fixes an issue or follows up on proposed modifications in a
          pull request, various commit messages refer to other resources such as
          issue tickets, i.e., \texttt{'Close \#306 and \#359'}, which is not a
          useful commit message on its own. However, this might be a predictor
          for fix commits. Therefore, we add a feature that is equal to 1.0 if
          the commit message contains a reference by means of a hashtag followed
          by digits, and 0.0 if this is not the case. Next to git issues, the
          Jira tool\footnote{\url{https://www.atlassian.com/software/jira}} is
          often used as a bug and issue tracking system for OSS projects. Jira
          references match the pattern \texttt{NAME-REF}, e.g., for Apache Solr a
          Jira issue can be \texttt{SOLR-12345}. Therefore, we add another
          feature that is equal to 1.0 if the commit message contains a Jira
          reference found through this pattern, and 0.0 if this is not the case.

    \item[Path similarity score.]\label{rvc_path_similarity} It can be the case
          that a vulnerability description provides a lot of information, and
          can even mention the file that was vulnerable. When such a file
          (usually a file path) is provided, it is very likely that the fix
          commit should change this file. Therefore, we add a path similarity
          feature. First, we extract the paths from vulnerability descriptions.
          When a path is found, we split this path and all paths the commit
          changes on dots and slashes, and reverse the list of path components.
          Depending on whether or not the path extracted from the description
          contains an extension, we remove the extensions from these lists. For
          every file we count how long the longest sub-list of similar elements
          is when starting from the front. Thus if \texttt{'example/file.py'} is
          extracted from the description and one of the changed files was
          \texttt{'project/main/example/file.py'} the longest matching sub-list
          is of length three (\texttt{['py', 'file', 'example']}). If the found
          path contains an extension and the longest matching sub-list has
          length equal to 1, hence just a matching extension, we do not add this
          score. We compute this score for every changed file and add the
          maximum score to the feature, which we do for all paths found in the
          vulnerability description --- remember that a description may mention
          multiple paths. The pseudo-code of how the value for this feature is
          computed is detailed in \Cref{algo_path_similarity}.

          \begin{figure}[ht]
              \begin{minipage}{0.75\textwidth}
                  \begin{algorithm}[H]
                      \caption{Compute path similarity
                      score}\label{algo_path_similarity}
                      \begin{algorithmic}[1]
                          \State path similarity score = 0; \For{path in the
                          vulnerability description:} \State intermediate score
                          = 0; \State Split path on special characters;
                          \For{changed file in all changed files:} \State Split
                          changed file on special characters; \State similarity
                          score = longest matching sub-list from the end;
                          \If{path contains extension \textbf{and} similarity
                          score $>$ 1:} \If{similarity score $>$ intermediate
                          score:} \State intermediate score = similarity score;
                          \EndIf \EndIf \EndFor \State path similarity score +=
                          intermediate score; \EndFor \State \textbf{return}
                          path similarity score;
                      \end{algorithmic}
                  \end{algorithm}
              \end{minipage}
          \end{figure}

    \item[Lexical similarity features.]\label{rvc_lexical_similarity} While
          other approaches try to predict whether a commit is a fix commit for a
          vulnerability, our approach is centered around predicting whether a
          commit is a fix for a \emph{specific} vulnerability. Therefore, we
          assume that the actual fix commits have high lexical similarity with
          the vulnerability description, through e.g., mentioning the same
          vulnerable components and vulnerability threats. The abstract approach
          to model such situation lies in mapping a vulnerability description
          and the commit information onto a joint feature space where their
          similarity can be established. A widely adopted method for this is
          term frequency-inverse document frequency (TF-IDF), which has been
          applied to vulnerability descriptions
          before~\cite{sawadogo_learning_2020}. TF-IDF increases importance of
          words that do not occur often in the entire corpus, and thereby
          adjusts for words that appear more frequently in the entire corpus. As
          the lines in the git diff contain code, the standard code constructs
          are becoming less important that project specific component names such
          as class and function names. For this, the \texttt{TfidfVectorizer}
          function from Scikit Learn's feature extraction module was
          used~\cite{scikit-learn}.

          We perform the same pre-processing steps on the vulnerability
          description and on the commit content. However, sometimes the
          \texttt{git diff} contains hundreds of thousands lines. Therefore, we
          consider only the first ten-thousand lines. For each advisory record,
          we create a separate array for each commit component; Thus, one corpus
          contains all commit messages of all candidate commits, one all changed
          file names, and one all \texttt{git diff} information.

          We then compare the elements in these arrays with the vulnerability
          description, code tokens, and the advisory references content. The
          code tokens are the tokens in the vulnerability description that are
          in CamelCase, snake\_case, or dot.case, and are no path. The advisory
          references content was extracted through scraping all references the
          NVD refers to, from which we take the 20 most frequently occurring
          words after the texts have been preprocessed. To the vulnerability
          description that we compare with the commit messages, we add a list of
          fix indicating words in order to assign them additional importance:
          \texttt{['security', 'cve', 'patch', 'vulnerability', 'vulnerable',
          'advisory', 'attack', 'exploit', 'exploitable']}.

          Next, we fit three different TF-IDF vectorizers on the three arrays.
          The vectorizer transforms the description, code token list, advisory
          references content, and the commit content to vectors. We compute the
          cosine similarity for each candidate's feature vector with the
          description vector, the references content vector, and the code token
          vector. The cosine similarity score is a value that is between 0.0 and
          1.0, where the higher the score the higher the similarity. This
          results in three vectors for each commit feature, hence nine lexical
          similarity features.

    \item[Commit distance to vulnerability release.]\label{rvc_commit_distance}
          We also add two features reflecting the distance between the
          vulnerability release date and the commit date, one reflecting the
          distance if the commit was done before the vulnerability release, and
          one if the commit was done after the vulnerability release.  We
          decided to create two distinct features to allow for an easier
          interpretation of the scores that are assigned to the features. The
          value for the feature reflecting the time distance before the
          vulnerability release is equal to 0.0 if the commit is after the
          release date, and between 0.5 and 1.0 if the commit was before the
          vulnerability release. The closer the commit is to the release, the
          higher the value. Hence, the 'first' candidate commit has a value of
          0.5 and the 'last' commit that was still before the vulnerability
          release has a score of 1.0 for this feature. For the time distance
          after the vulnerability release, this is the other way around. If a
          commit was before the vulnerability release date the score is 0.0, and
          if the commit was after the score is between 0.5 and 1.0, where the
          higher the score the closer the commit was to the vulnerability date.
          Please consider that 'time distance before' and 'time distance after'
          could be combined into one feature like 'absolute time distance'.
          However, we decided not to go in this direction for the sake of
          interpretability.

    \item[Reachability score.] It can take a long time before a vulnerability is
          released after it has been actually discovered and patched (
          \Cref{fig:selection_visualized}), whereby the two time-based features
          might not be a good predictor. As commits in GitHub are in a tree
          structure, not every commit is reachable from a given tag. Therefore,
          we add another feature that reflects the distance to a commit from the
          last vulnerable version. \Cref{fig:branch_based_versioning} presents
          an example of such tree structure and shows how the branches of this
          tree are used in software development, where commits are illustrated
          by circles and the tags are illustrated squares showing version
          numbers. Referring to this example, the commits that are reachable
          from e.g., version 1.6.0 are the commits that are before this tag and
          in the same branch (i.e., commits belonging to tag 1.5.0). Commits that
          are in the same branch but after the current tag (i.e., version 1.7.0)
          or in a different branch (i.e., version 1.4.2) are not reachable. This
          reachability can be translated into a predictor for being the fix when
          a vulnerability description mentions that e.g., version 1.6.0 is no
          longer vulnerable, the fix should be reachable from this tag.

          \begin{figure}
              \centering
              \fbox{
                  \includegraphics[width=0.95\textwidth]{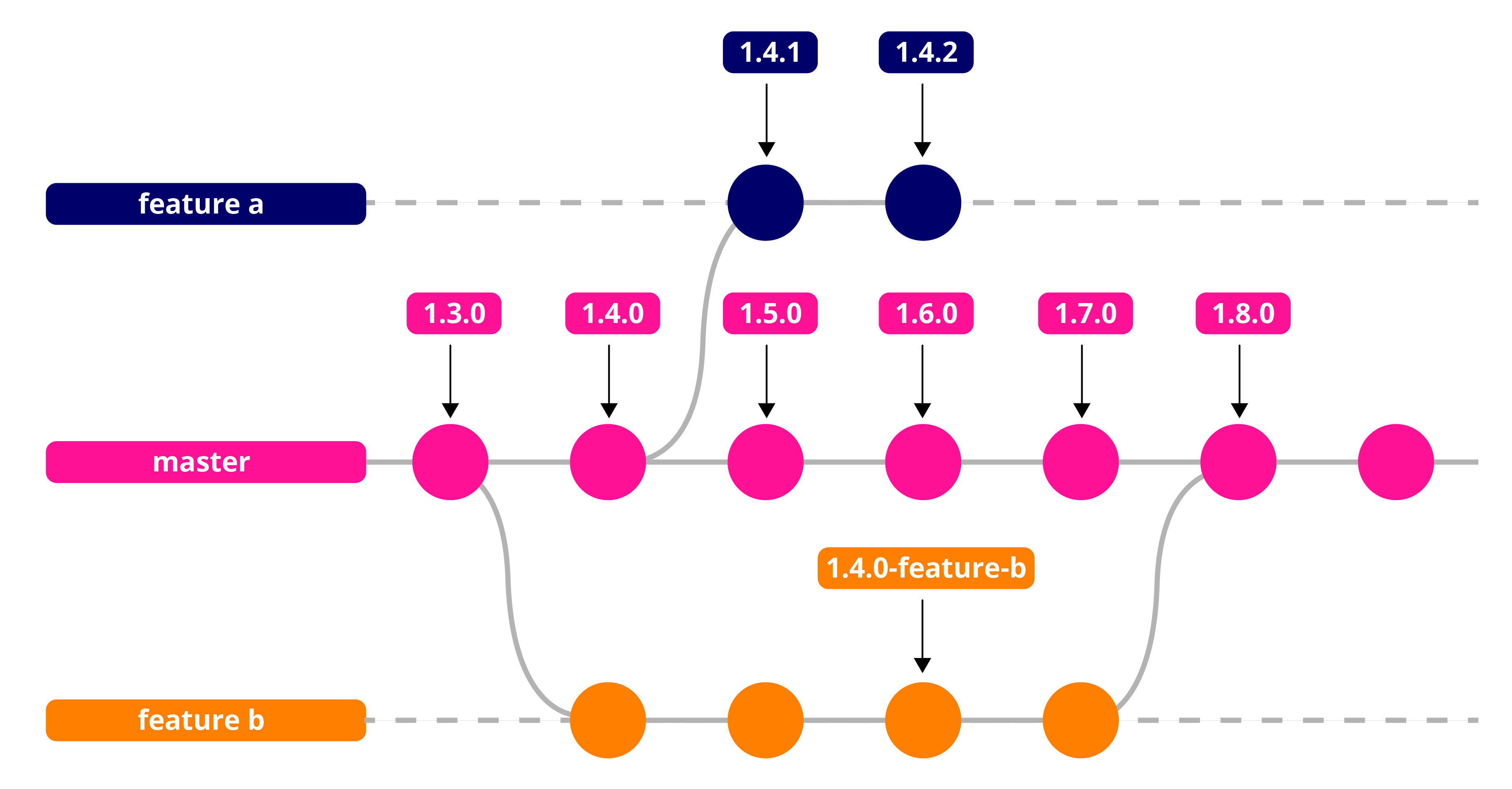}
              } \caption{An illustration of how Git's branch-based version works
              for an example with three different branches.}
              \label{fig:branch_based_versioning}
          \end{figure}

          We compute the reachability score by the following steps. First, we
          add all tags from the repository to a tree where only the tags from
          the same major and minor version are in the same branch. We sort the
          tags in the branches on the tag timestamp (visualized in
          \Cref{fig:tags_to_tags_tree} in \Cref{apx:tags_to_tree}). We then
          extract the version numbers from the vulnerability description, and
          map these to tags from the repository. For every tag that we find in
          the repository, we find the next tag by means of the sorted tags tree
          if there is a later tag with the same major, otherwise the original
          tag is used to calculate reachability from. For every candidate
          commit, we compute a list of scores where every element is the
          reachability score from the tags that have been found: this
          reachability score is 0.0 if it is not reachable (or not within 100
          days of the tag), and 1.0 - $x$ where $x$ is 0.01 for every day
          between the tag timestamp and the commit timestamp (i.e., 0.95 if the
          commit was 5 days before the tag has been created). For every
          candidate commit, we use the maximum value in this list as the value
          for the reachability feature.

          The one-hundred days difference we mentioned is  heuristic based on
          the data of~\cite{ponta_manually-curated_2019}, from which we could
          observe that most fix commits fall within a one-hundred days interval
          before or after the date of a tag (release). Therefore, we decided to
          provide a higher score for reachable commits within one-hundred days
          from a version that was mentioned. The commits that are not within
          one-hundred days but are reachable get the same score as commits that
          are not reachable. We take this approach as branches with new features
          are frequently merged with the main branch, whereby all these commits
          become reachable.  This can be seen in the example in
          \Cref{fig:branch_based_versioning}, where the 'feature b' branch is
          merged before tag '1.8.0' and the commits in this branch become
          reachable from '1.8.0', whereby almost all commits from the repository
          have become reachable. Therefore, if we were to assign a score for all
          reachable commits, the importance of this score might decrease as a
          large number of commits are reachable.

    \item[Vulnerability timestamp.] The last feature we add is the vulnerability
          timestamp to allow models to learn a certain development over time, as
          it can be the case that over time the OSS community has become e.g.,
          faster at fixing vulnerabilities.

\end{description}

\subsection{Feature scales and normalization}\label{experiment_scaling} When the
features in the input data have different scales, one feature can have more
influence than another feature. Therefore, it is advisable to scale features
before providing the data to a ML algorithm. The majority of the features is
already a value between 0.0 and 1.0 but for instance the number of hunks is an
integer that can be very large. To account for this, we apply the
\texttt{MinMaxScaler} from Scikit Learn~\cite{scikit-learn}. We could apply a
vulnerability specific scaling, where we fit-transform the \texttt{MinMaxScaler}
for every vulnerability on just the candidate commits for this vulnerability.
Hereby, e.g., the largest number of changed files from this candidate commit
selection could be ten, would be scaled to 1.0. For another vulnerability,
this number could be one-hundred, which would also be scaled to 1.0. For some
features this makes sense, such as the lexical similarity features as it might
be more valuable to know which of the candidate commits had the highest lexical
similarity with the vulnerability description than knowing that it had a
relatively high score regarding the lexical similarity for all candidate commits
and their vulnerabilities. However, for some features such as the number of
changed files, it might make more sense to look \emph{universally} (all
candidates for all vulnerabilities).

Therefore, we take a combination of the two: The features \textit{vulnerability
    ID in message, other vulnerability ID in message, referred to by NVD, referred
    to by advisories, message contains GitHub issue reference, message contains a
    reference to a Jira issue, time distance before,} and \textit{time distance}
after are already values between 0.0 and 1.0. We scale the features
\textit{number of hunks, average hunk size, (vulnerability timestamp,)} and
\textit{number of changed files} universally, as we regard these features not
vulnerability or project specific features. We scale the features \textit{path
    similarity score} and the nine \textit{lexical similarity features}
vulnerability specific, as we regard these features to be vulnerability specific
as they depend on the vulnerability description.

\subsection{Model Training}

After we create and scale the feature vectors for the candidate commits, the
candidate commits can be ranked based on the scores for these features. This
task can be approached as a ML task, by means of predicting the
probability of being a fix commit for this vulnerability and sorting the
candidates based on this probability score, for which we use the Scikit-Learn
\texttt{predict\_proba} method~\cite{scikit-learn}. Our choice of algorithms
and the report of their performance are presented in the next section.

\section{Empirical Evaluation}\label{evaluation}




\subsection{Data Preparation}

In order to train a ML model capable of predicting the probability of each
candidate to be a fix commit, train and test dataset need to be created.
This does not just yield randomly shuffling the vulnerabilities and selecting
a subset because of the imbalance of the dataset, as there can be thousands of
candidate commits for just one known fix commit. One of the decisions to make is
the number of negative samples for every positive instance~\cite{sabetta_practical_2018}.
In our research, we use a train and test dataset containing one negative and one
positive candidate commit for each vulnerability (from the same repository). We
then use this dataset was to train a classifier in classifying feature vectors
to either negative or positive. When ranking the candidate commits, we use this
classifier to predict for every feature vector the probability of being a fix
commit, so we can sort the candidate commits based on this score.

\subsection{Experiment Setup}

The algorithms we evaluate are Support Vector Machine Classifier (SVC), Logistic
Regression (LR), k-Nearest Neighbors Classifier (KNNC), Random Forest (RF), and
Adaboost Classifier (AB). For each algorithm, we tune the parameters by means of
bayesian optimization (\Cref{tab:bayesian_optimization_parameters}). Such
method is an effective way to optimize functions such as the hyperparameter
setting, as the sampling of parameter values does not occur randomly but an
acquisition function is used to estimate the likelihood of a sample being worth
to evaluate based on uncertainty. The sample that has the highest value for this
expected improvement is evaluated, reducing uncertainty in this area. After
twenty iterations, the optimal model is returned and it is subsequently used to
predict for every feature vector its probability of corresponding to a fix
commit. The candidate fix commits are sorted over such probability score.
Thereby, the candidate with the highest probability of being a fix commit
results to be in the first position. We do this ten times with randomly selected
train and test splits, and by using the models from
Scikit-Learn~\cite{scikit-learn}.

\begin{table}
    \centering
    \footnotesize
    \caption{The models and their parameters tuned using Bayesian optimization.}
    \begin{tabular}{l|ll}
    \toprule
         \textbf{Algorithm}      & \textbf{Parameter (Range (scale)) }
         & \textbf{Parameter (Range (scale))} \\
    \hline
    \midrule
         SVC            & C (1e-12 - 1e+12 (log))       & gamma (1e-12 - 1e+12
         (log))       \\
         LR             & C (1e-12 - 1e+12 (log))       & l1-ratio (0 - 1
         (real))           \\   
         KNNC           & n\_neighbors (1 - 200 (int))  & p (1 - 2 (int))
         \\
         RF             & n\_estimators (1 - 400 (int)) & max\_depth (1 - 20
         (int))         \\
         AB             & n\_estimators (1 - 400 (int)) & learning\_rate (1e-6 -
         2 (real))  \\
    \bottomrule
    \end{tabular}
    \label{tab:bayesian_optimization_parameters}
\end{table}

The selection based on the vulnerability release date results in
\candidatesinselection{} candidate commits for \nvulnerabilitiesindata{}
vulnerabilities, reducing the number of commits to consider to
\filterpercentage{}\%. From these \candidatesinselection{} candidate commits, we
removed \candidatesnotrelevant{} commits as they do not change any file with a
relevant extension, which reduced the number of candidate commits to
\totalfilteringpercentage{}\%.

For every candidate commit in the filtered subset, we created a feature vector
composed of \nevalcomponents features. These feature vectors are then used to
predict the probability of a candidate commit being a fix commit for a given
vulnerability. The results in this section are for the
\nvulnerabilitiesinselection vulnerabilities for which the selection includes at
least one known fix commit (making it possible to rank). The results have been
calculated on the same ten randomly created splits of the data, with 933
vulnerabilities in the training sets and 233 vulnerabilities in the test set
(80/20 train/test split). We compare the performance of \nmlmodelsevaluated
different models to the baseline methods described in \Cref{previous_methods}: (i)
CVE ID in message, (ii) referred to by the NVD, and (iii)
\cite{jimenez_enabling_2018}. To analyse the extent to which the NVD refers to
fix commits for the vulnerabilities in our dataset, we use the NVD CVE data
feeds\footnote{\url{https://nvd.nist.gov/vuln/data-feeds}} to extract the URLs
that can be found under the section 'References to Advisories, Solutions, and
Tools' on the NVD-page of the vulnerability.

\begin{table}
    \centering

    \caption{The performance of the different machine learning algorithms to
    find fix commits compared to the baseline methods (i) CVE ID in message,
    (ii) Referred to by the NVD, and (iii) \cite{jimenez_enabling_2018}. For
    each model, the table shows the mean recall achieved when recommending 1, 5,
    10, or 20 commits and the average position of the fix commits in the
    recommended ranking. The best result is depicted in bold, while the standard
    deviation is denoted between parentheses.}

    \resizebox{\textwidth}{!}{
        \begin{tabular}{l|llllll}
            & \textbf{Method} & \textbf{Top-1 Recall} & \textbf{Top-5
            Recall} & \textbf{Top-10 Recall} & \textbf{Top-20 Recall} & \textbf{Avg.
            Pos.}\\
            \hline
            \midrule
            \prospector & Adaboost & 62.67 (2.5) & 75.89 (2.81) & 83.26 (2.13) &87.68
            (2.1) & 24.25 (8.86)\\
            & k-Nearest Neighbors & 18.04 (9.38) & 27.13 (15.24) & 39.22 (14.86) &
            54.70 (12.55) & 108.65 (35.5)\\
            & Logistic Regression & \textbf{65.06 (1.91)}& \textbf{77.68 (1.41)} &
            \textbf{84.03 (1.22)} & \textbf{88.24 (1.0)} & 22.89 (5.47)\\
            & Random Forest & 56.77 (1.96) & 73.13 (1.39) & 81.54 (1.58) & 87.58
            (1.81) & \textbf{18.50 (4.81)}\\
            & Support Vector Classifier & 61.89 (5.44)& 75.72 (3.65) & 82.64 (2.65)
            & 87.37 (2.31) & 22.48 (5.83)\\
            \hline
            Baseline & CVE ID in message & 12.60 (1.93) & & & & \\
            & Referred to by the NVD& 12.04 (2.46) & & & & \\
            & \cite{jimenez_enabling_2018} & 23.24 (2.45) & & & & \\
            \bottomrule
        \end{tabular}}
    \label{tab:ranking_results_extended_dataset}
\end{table}

\subsection{Results of the Model Comparison}

\Cref{tab:ranking_results_extended_dataset} shows the results. For
\rankingprecision{}\% of the vulnerabilities a fix commit is the first result,
and for \rankingrecallten{}\% at least one fix is in the top ten (using the
best-performing model, i.e., logistic regression).

\prospector outperforms all the considered baselines. Indeed, when looking at our
data set, we find that for only 12.60\% of the vulnerabilities, at least one fix
commit can be found by looking for the CVE ID in the commit messages. By
counting how often the first eight characters of the fix commit ID could be
found in one of the NVD references, we find that at least one fix commit can be
found for only 12.04\% of the vulnerabilities in our data. Through applying the
approach of Jimenez et al.~\cite{jimenez_enabling_2018}, we find that for only
\vulnerabilitiesfoundjimenez vulnerabilities \precisionjimenez{}\% a fix can be
found through looking for the vulnerability identifier in the commit message or
following the references in the NVD. This result emphasizes the need for a more
reliable method to find security-relevant commits, such as \prospector. Please
consider that all machine learning models but kNN outperform the baselines,
suggesting that the feature engineering previously described can significantly
improve the performance of the recommendation task.

\subsection{Experimental Research Design}
With this research, we propose an approach to automate the process of finding
fix commits for vulnerabilities in open-source software. As the vulnerability
information that can be found on vulnerability knowledge advisories as the NVD
contain is not machine readable and needs to be extracted and processed before it
can be used for such an automation task, the research was divided in three
steps: extract, filter, and rank. In the extract phase, the focus was on
extracting valuable information that should be incorporated in the advisory
record. In the filtering phase this advisory record was used to come to a
selection of candidate commits. In the ranking phase, these candidate commits
were ranked based on the probability of being a fix commit for the provided
advisory record. To answer the main research question, we created three
sub-questions each addressing one of the pillars of this research to test our
approach:

\begin{center}
    \textbf{How can automation be used to reduce the effort needed in
        finding fix commits\\ for known vulnerabilities in open-source software?}
\end{center}

Stemming from this general research question, three more concrete research
questions emerge, namely:
\begin{enumerate}
    \item\label{rq_1} What vulnerability information should be extracted from
          resources as the NVD and used in the advisory record of a vulnerability?

    \item\label{rq_2} How can the large number of commits in repositories be
          reduced to a subset of candidate commits through filtering out irrelevant
          commits?

    \item\label{rq_3} How to use ML to rank the candidate
          commits with the fix commits as high as possible?
\end{enumerate}

To answer these questions, we examine the coefficients of the logistic
regression model (i.e., the one providing the best predictive performance),
which can be found in \Cref{tab:lr_coefficients}. The higher the absolute value
of a coefficient, the more this feature is important for predicting.
Furthermore, a positive value means that a feature positively contributes to the
model, while a negative value implies a negative contribution. In other words,
if a feature is positive/negative, a high positive/negative instance value leads
to predict a certain commit as a fix.

\begin{table}
    \caption{Coefficients for the logistic regression model.}
    \centering
    \footnotesize
    \begin{tabular}{l|r}
    \toprule
    \textbf{Feature}                            & \textbf{Coefficient}  \\
    \midrule
    \hline
    Time distance before                        & 8.189  \\
    Time distance after                         & 8.095  \\
    Vulnerability ID in message                 & 7.363  \\
    Referred to by advisories                   & 5.054  \\
    Git diff similarity score                   & 4.412  \\
    Message similarity score                    & 3.244  \\
    Changed files similarity score              & 2.831  \\
    Referred to by NVD                          & 2.559  \\
    Reachability score                          & 2.462  \\
    Other vulnerability ID in message           & -1.994 \\
    Message similarity score code tokens        & -1.165 \\
    Message similarity references content       & 1.035  \\
    Vulnerability timestamp                     & 0.969  \\
    Number of changed files                     & 0.526  \\
    Path similarity score                       & -0.483 \\
    Message contains Jira reference             & 0.283  \\
    Git diff similarity score code tokens       & -0.274 \\
    Number of hunks                             & -0.111 \\
    Git diff similarity references content      & -0.110 \\
    Average hunk size                           & -0.100 \\
    Changed files similarity references content & -0.097 \\
    Changed files similarity score code tokens  & 0.090  \\
    Message contains Git issue reference        & -0.047 \\
    \bottomrule
    \end{tabular}
    \label{tab:lr_coefficients}
\end{table}

\textbf{RQ1}: The advisory records contain vulnerability information that was
extracted from resources as the NVD and translated into features. The
vulnerability release date was used to select a subset of candidate commits and
to calculate the features \emph{time distance before} and \emph{time distance
after}. These two features have the highest positive coefficient and are
essential in the ranking procedure. The closer the commit is to the
vulnerability publication date, the more likely it is to be a fix commit. Please
consider that if the timestamp differs more than two years from the actual date
the vulnerability has been published, the fix commit is not even considered a
candidate.

The vulnerability description is used for six of the nine lexical similarity
features, the path similarity component, and the reachability score component.
This reachability score is calculated through extracting the affected versions
and calculating the reachability for the commits, and the path similarity score
is computed after extracting the paths that are mentioned in the description and
calculating the overlap with the paths that are changed by the commit. All of
these features have a relatively high coefficient value, except for the path
similarity score, whereby the path similarity score does not appear to be a good
predictor. The references that are in the advisory record (to which the NVD
refers) are used for three lexical similarity features, and for the two
'referred to by' features. The two referred to features are of high predictive
value (as the coefficients are high), so the method of extracting all URLs that
are on these sites was a method with a positive effect on the prediction
capabilities of the model. Hereby, this has proven to be a relevant addition to
the advisory records. The lexical similarity features however have a low or even
slightly negative coefficient.

\textbf{RQ2}: As the total number of candidate commits would equal
\totalnumberofcommits{} without filtering, the filtering method used in our
implementation reduced this large number to only \filterpercentage{}\% while
maintaining a selection recall of \selectionrecall{}\%. After the selection was
reduced to \filterpercentage{}\% of the total number of commits, another
\percentageirrelevantextension{} could be discarded through removing commits
that do not change a relevant extension. From these \candidatesnotrelevant{}
candidate commits, no commit was known as a fix commit, whereby this was a very
effective filtering step. The two steps combined reduced the total number of
commits to only \totalfilteringpercentage{}\%. Therefore, this filtering
procedure was an effective application of our approach, and is discussed more in
depth in \Cref{discussion}.

\textbf{RQ3}: The method of creating feature vectors composed of several
features that contain vulnerability specific scores has proven to be very
effective, as our implementation ranked a fix in the top-10 for
\rankingrecallten{}\% of the vulnerabilities.

\section{Discussion}\label{discussion}

\subsection{Dataset}
\label{dataset_discussion}

Where other research focused on predicting whether a commit is a fix commit or
not in a balanced humanly curated dataset, our research was looking for fix
commits in real repositories. Where Sabetta et al.~\cite{sabetta_practical_2018}
worked with 456 positive (fixes) and 2,259 negative samples and Sawadogo et
al.~\cite{sawadogo_learning_2020} with 2,879 positive and 6,368 negative
samples, our research was conducted on a dataset containing
\nknownfixcommitsindata{} positive samples and 25,017,005 negative samples.
Moreover, their datasets are divided into fix commits for sure and commits that
are for sure not fix commit. Our dataset does not have this clear distinction
for two reasons. First, there can be more unknown fix commits. When examining
the results of \prospector, it was found that for some vulnerabilities
for which the fix commits were not the first ones, other unknown fix commits
were at the top, whereby the precision score was falsely reduced. Second, there
might be over a hundred-thousand fix commits among this large number of commits,
fixing other vulnerabilities than the ones that are queried. Finally, we did not
conduct our study on the dataset provided by Li and Paxson~\cite{li2017large}
for a two-fold reason. On the one hand, its quality was not manually validated.
On the other hand, although the authors searched for fixes of 80,741 CVEs, they
could find only 3.094 commits. Therefore, we deem the commits in that dataset
not representative of the overall population of fix-commits.

Where the other research was centered around creating statistics that tell
something about the probability of a commit being a fix commit, this method does
not work for our use case as we are looking for specific fixes for specific
vulnerabilities. When looking at the coefficients for the commit statistics
features (\Cref{tab:lr_coefficients}), it can be seen that the importance
assigned to these features is very small. This can be explained by the fact that
there are fix commits in the data with a negative label, as they are not the
fixes that were looked for. Moreover, as some vulnerabilities belong to the same
repository, the same commit may be a positive sample for one vulnerability, and
a negative sample for another. While the total number of candidate commits
equals \totalnumberofcommits{}, these candidate commits consist of 906,419
commits. Therefore, using a database was not only beneficial during (iterative)
development but also means that there needed to be 67\% fewer commit extraction
operations. Please consider that although we did not explicitly consider lines
of code, these features are in a way correlated with this measure (i.e., number
of hunks, average hunk size, and number of changed files).

It is important to observe that there is no one-to-one correspondence between
vulnerabilities and commits: a single commit can fix multiple vulnerabilities,
and a vulnerability might be addressed by a set commits that, together,
represent the fix. Also, the dataset used for our experiments has been manually
curated and it may be biased by the purposes for which it was created (that is,
to feed the software composition analysis tool \textsc{Eclipse Steady}. As a
result, for a given vulnerability there might exist more fixes than captured in
the dataset (e.g., coming from other branches than the ones considered when
gathering the data).

\begin{table}
    \centering

    \caption{The performance of the different machine learning algorithms to
    find fix commits compared to the baseline methods (i) CVE ID in message,
    (ii) Referred to by the NVD, and (iii) \cite{jimenez_enabling_2018} on the
    publicly available part of our dataset. For each model, the table shows the
    mean recall achieved when recommending 1, 5, 10, or 20 commits and the
    average position of the fix commits in the recommended ranking. The best
    result is depicted in bold, while the standard deviation is denoted between
    parentheses.}

    \resizebox{\textwidth}{!}{
        \begin{tabular}{l|llllll}
            & \textbf{Method} & \textbf{Top-1 Recall} & \textbf{Top-5
            Recall} & \textbf{Top-10 Recall} & \textbf{Top-20 Recall} &
            \textbf{Avg. Pos.}\\
            \hline
            \midrule
            \prospector & Adaboost & 47.85 (8.87) & 65.37 (6.78) & 74.44 (5.76) &
            82.46 (4.53) & 35.59 (15.09) \\
            & k-Nearest Neighbors & 10.38 (1.47) & 12.86 (2.68) & 21.00 (2.74)&
            37.16 (3.18) & 166.23 (37)\\
            & Logistic Regression & \textbf{55.37 (3.92)} & \textbf{70.78
            (3.81)} & \textbf{78.04 (4.08)}& \textbf{84.97 (2.15)} & 34.37
            (16.17)\\
            & Random Forest & 45.31 (3.04) & 62.78 (4.58) & 73.74 (4.08)& 82.37
            (3.37) & \textbf{25.85 (11.54)}\\
            & Support Vector Classifier & 46.35 (17.17) & 59.60 (20.5) & 68.91
            (19.04) & 77.09 (17.29) & 50.78 (52.39)\\
            \hline
            Baseline & CVE ID in message& 6.49 (6.49) & & & & \\
            & Referred to by the NVD & 9.95 (1.59) & & & & \\
            & \cite{jimenez_enabling_2018} & 16.44 (2.50)& & & & \\
            \bottomrule
        \end{tabular}}

    \label{tab:ranking_publicly_available_dataset}
\end{table}

Please consider that only part of our dataset is publicly available. For the
sake of replicability, we repeated our analysis on its publicly available part
composed of 1,181 fix commits for 614 vulnerabilities. As in the complete
experiment, reported in \Cref{evaluation}, we applied an 80/20 training/testing split,
having 123 vulnerabilities in the test split of which on average 115.7 CVEs with
at least one known fix commit in the selection. The results, shown in
\Cref{tab:ranking_publicly_available_dataset} show results similar to those
previously analyzed and reported in \Cref{tab:ranking_results_extended_dataset}.
The slightly inferior performance can be justified by the smaller training set.

\subsection{Filtering}
We estimated that an interval of two years before and one-hundred days after the
CVE publication date would yield a good interval. However, for repositories that
have a large number of commits per year, the number of candidates can be very
large. We calculated that selecting all commits within two years before and one
hundred days after the CVE publication would result in a total of 4,972,699
commits for all vulnerabilities, and found that this selection would yield a
selection recall of 95.35\%, meaning that for 95.35\% of the vulnerabilities at
least one known fix commit would be in the selection.

Another interval to select candidates could be filter out commits that have been
committed too far from the CVE publication date by means of number of commits,
e.g., it would be unlikely that there have been ten-thousand commits after this
commit before the vulnerability was published. We decided that this number
should result in a selection with recall similar to the selection by means of
the number of days, which could be achieved through selecting five-thousand
commits before and one-hundred commits after the CVE publication date. This
interval would yield 4,652,884 candidates with a selection recall of
96.07\%\footnote{Note that they are not 6.364.800 commits (i.e., 5100 commits
    for all 1248 vulnerabilities) because some repositories have fewer than 5000
    commits before the CVE publication date.}. As this was still a large number of
commits, we explored a combination of the two: Selecting all commits within two
years before and one hundred days after the release date with a maximum number
of commits. We found that this combination of the two, selecting all commits
within two years before and one hundred days after the release date with a
maximum of respectively 5215 and 100 commits, would result in a recall 93.59\%
but would require 2,753,058 commits. As this selection had less than three
percentage lower recall but forty percent less commits, we took this method as a
filtering option.

This selection strategy is visualized in \Cref{fig:selection_visualized}.
An overview of the strategies can be found in 
\Cref{tab:nvd_release_based_sampling}. These numbers have been calculated without
accounting for duplicate commits when vulnerabilities belong to the same
repository and the commits only need to be added to the database once.

\begin{table}
    \centering
    \footnotesize
    \caption{The effect of different selection strategies to select candidate commits based on their distance from the vulnerability publication date on the number of candidate commits and the selection recall. The selection recall is the percentage of vulnerabilities for which at least one fix commit can be found in the selection.}
    \begin{tabular}{l|ll}
    \toprule
    \textbf{Method}                  & \textbf{Selection recall \%}          & \textbf{n candidate commits}  \\
    \hline
    \midrule
    Time based selection    & 95.35              & 4,972,699 \\
    Commit based selection  & 96.07              & 4,652,884 \\
    Combination             & 93.59              & 2,753,058 \\
    \bottomrule
    \end{tabular}
    \label{tab:nvd_release_based_sampling}
\end{table}

As the scope of our research was to explore whether it is possible to find the
fix commits in a fully automated manner, this range of two-years before and with
a maximum number of 5215 commits, and one-hundred days after with a maximum of
100 commits was taken. When specifying the bounds to be for instance
one-thousand commits before the release date and one-hundred after the release
date, the selection recall would be 85.82\%. When using our approach, one could
for instance start with a smaller window and widen this window to see what the
best interval is.

In our implementation, another filtering method was evaluated as well.
As vulnerability descriptions often mention the software versions that are
affected, we assumed that this information could be used to come to a selection
of candidate commits. However, when working with a subset of 780 vulnerabilities
that at least mentioned one version with \textit{digit.digit}, recall could not
get higher than 67\%.
We also found that a more sophisticated method yielded lower recall: this was
the case when using words as \texttt{'through'} or \texttt{'including'} to
determine whether a given version was the last vulnerable or the first fixed.
When analysing manually the cases where this approach would fail, we found that
there were too many inconsistencies in the tags in the OSS source repository.
Even when considering the entire interval of the first tag of the previous minor
version and the last tag of the next minor version (i.e., 4.3.2 is mapped onto
4.2.0 and 4.4.9), recall would not increase. While more and more effort was put
in pre-processing tags to account for the lack of semantic versioning, the
results on the test set would not improve. While we  expected to have a recall
of nearly 100\% (as every version mentioned in the description was extracted and
this was mapped onto a wide interval), only for 67 out of 100 vulnerabilities in
an unseen test set a fix commit could be found in the interval. An examination
of the 33 instances for which the mapping failed showed that 22 description
mentioned the wrong version (which was neither the last vulnerable nor the first
fixed), 7 versions mentioned did not match any tag in the repository, 4 fix
commits did not have a tag (which is rather a problem of the dataset we used as
training data).

This inconsistency in vulnerability descriptions was also found by Dong et
al.~\cite{dong_towards_2019}, who found that only 58.62\% of the CVE summaries
have exactly the same vulnerable software versions as those of the of the
standardized NVD entries. According to their case study, out of 185 versions
that were stated to be vulnerable only 64 were actually vulnerable. They found
that one of the causes was typos in the advisory, where, e.g., 0.22 was
mentioned instead of the actual vulnerable version 0.2.2, or 0.8.6 instead of
0.6.8.

\subsection{Ranking}
When comparing our ranking score to the method of Jimenez et
al.~\cite{jimenez_enabling_2018}, who would be able to find a fix for
\vulnerabilitiesfoundjimenez{} of the vulnerabilities in this dataset and
thereby resulting in a precision score of just \precisionjimenez{}\%, our
implementation has a precision of \rankingprecision{}\%. When used in practice,
\prospector could perform even better: when examining the output of the
tool, we found that some fix commits that were not known and thus labelled as a
non-fix in the reference training data could actually be found by the tool,
reducing the evaluation score.

Not all features that were engineered proved to be valuable. We  expected a path
being mentioned to be a very good indicator for a commit being a fix commit,
however the coefficient of the LR model is negative. From 226 of the 1,248
vulnerabilities a path was extracted, which might be too few to be of great
importance in general. It is also possible that there are too many commits that
change the paths that are mentioned, therefore this component might not contain
relevant information to distinguish among them.

Vulnerability descriptions from the NVD do not always provide enough information
such as code constructs whilst the NVD provides a lot of references. Therefore,
we assumed that the performance can be improved through extraction textual
information from the advisory references. However, as the coefficients of the
lexical similarity with the references content indicate, extracting the twenty
most occurring words from all references might not be the best method for
extracting relevant information from the sites to which the NVD refers. An
explanation for this could be that often some of the references are from the
same website, whereby the context words that are on this site might end up among
the twenty most occurring words.

The \textit{message contains git issue reference} and \textit{message contains
    Jira references features} are two features that are not of prediction value in
the current method, as it just reflects whether a commit message refers to an
issue. From all candidate commits, 20.71\% of the commits refers to a git issue
and 53.80\% refers to a Jira issue. Therefore, this component does not hold any
information of the likeliness of being a fix for the queried vulnerability.
Extracting the content from this issue can be valuable, as this can be used for
computing the lexical similarity with the vulnerability descriptions. However,
as this would mean that the content needs to be extracted for 482,789 GitHub
Issue pages and 1,253,889 Jira tickets, we regarded this out of scope for this
research.

When computing the lexical similarity between the vulnerability description and
the commit message, we added a list of fix indicating words to emphasize the
importance of these words in commit messages. Among other words, security and
patch are in this list. However, more than thirty-thousand commits contain the
word security and over sixty-thousand commits contain the word patch. Hereby,
this might introduce noise instead of helping the prediction.

Moreover, the lexical similarity was computed through vectorizing the
descriptions using TF-IDF and computing lexical similarity. One could explore
the use of embeddings. \cite{han_learning_2017} tried to predict the severity
score of a vulnerability based on only the vulnerability description and found
that a shallow CNN trained on word embeddings performs significantly outperforms
the classification methods using TF/IDF features or only word embeddings, when
trained with as minimal one-thousand vulnerabilities. As the training set for
our research consists of 933 vulnerabilities, we decided to stick with the
TF-IDF method.

For 732 of the \nvulnerabilitiesindata{} vulnerabilities code tokens were
extracted from the description. We assumed that this would be a strong indicator
for being the fix commit, especially when comparing this with the git diff. That
the coefficients for these lexical similarity features are this low can be due
to various reasons. Firstly, it could be the case the TF-IDF combined with
cosine similarity is not the right method for this comparison e.g., due to size
of the git diff. Secondly, as the git diff can have a very large number of
lines, only the first ten-thousand lines were considered, therefore the relevant
lines might be ignored. Lastly, the output of the \texttt{git diff} command
shows  the actual lines of code that have been modified, and just a few context
lines. These context lines are the lines before and after the changed lines, and
this context size can be changed. In our implementation, the context size was
set to one (just one line above and one below the changed lines are in the
diff), but using larger values  might be beneficial.

\subsection{Scope of the Study and Threats to Validity}\label{limitations}

Our results were obtained on a subset of \nvulnerabilitiesinselection{}
vulnerabilities for which the selection yields at least one known fix commit.
Though this selection has \selectionrecall{}\% recall, this is a limitation of
the ranking results. Therefore, when taking into account that these results have
been calculated after the selection with \selectionrecall{}\% recall, the 'real'
ranking precision is \realrankingprecision{}\% and the ranking recall in the top
ten is \realrankingrecallten{}\%.

As it is known that there can be a significant delay between the time at which
the vulnerability is patched and the time at which a CVE is assigned, our
selection method might not be the best. When examining some vulnerabilities for
which the selection failed, we found that this is often due to a delay in the
vulnerability disclosure. For example, \textit{CVE-2020-28724} was published on
18 November 2020 but its fix commit was made on 14 April 2016.

When applying \prospector in practice, we observed that
\prospector could often find fix commits that would not have been
identified easily without the tool, as there were no advisory pages reporting
the commit. However, we did not compare the required effort for finding fix
commits with using our tool to the human baseline, whereby we cannot quantify
the decrease in required effort. This aspect will be the subject of future work.

Furthermore, we recognize that the filtering step could negatively impact the
recall of the approach. Researchers and practitioners should find the correct
balance between lower recall and higher scalability when filtering commits. In
our experiment, we selected all commits within two years before and one hundred
days after the release date. This selection strategy reduced the recall to
\selectionrecall{}\% but analyzing only \filterpercentage{}\% of commits.

\section{Related Work}\label{related_work}

The lack of information on vulnerabilities of OSS makes it hard to maintain a
secure software development supply chain. To account for this lack of
information, various researchers have spent their effort using machine learning
to analyze source code repositories.

Sabetta and Bezzi~\cite{sabetta_practical_2018} used an NLP-inspired method to
represent commits and to distinguish fix commits (positives) from non-fixes
(negatives). For each commit, bag-of-words was created by extracting the named
entities from the changed source code, such as the names of classes, functions,
parameters, variables. To come to a prediction, these documents were treated as
documents written in natural language which were classified using standard
document classification methods. The same problem was also tackled
by~\cite{Zhou2017} based on a larger dataset, using a set of models arranged in
a stacked configuration; however, they did not consider the content of the
commit diff (the code changes), but only the natural language content of the
commit message.

Where Sabetta and Bezzi~\cite{sabetta_practical_2018} used a SVM for the
classification task at hand, Lozoya et al.~\cite{CabreraLozoya2021} investigated
how to represent source code in a way that it can be used effectively for deep
learning and showed that transfer learning could be effectively used to cope
with the scarcity of labeled data.

Sawadogo et al.~\cite{sawadogo_learning_2020} used co-training of two
classifiers to automate the identification of source code changes that are
security patches. One of the classifiers uses text features based on the meta
data in the commit messages, and the other one uses code change details such as
the number of lines that have been modified. As they acknowledge themselves as
well, their research relies on scarce data, and they are assuming that the
labeled dataset is perfect and sufficient. However, during sampling of the
negative instances (commits that do not fix a vulnerability) they only sampled
commits that do not contain fix indicating words like "security" or
"vulnerability" in the commit logs, while one of their classifiers is predicting
whether a commit is a fix or not based on these commit logs. In the dataset
used in this research, only 0.59\% of the commits for which the message contains
'security' is a patch for one of the vulnerabilities in the dataset, which
means the dataset used in their research does not reflect the actual world at
all.

Previous research mainly focuses on security-relevant classification based on a
small, manually curated, balanced dataset. Ponta et
al.~\cite{ponta_manually-curated_2019} published a manually curated dataset
containing fixes for 624 publicly disclosed vulnerabilities. They reviewed
available information and searched for the corresponding fix commits in the code
repository, resulting in 862 unique fixes. Please consider that in this paper,
we used an extended version of the same dataset that includes data of additional
vulnerabilities not covered in ~\cite{ponta_manually-curated_2019}.

The process of Ponta et al.~\cite{ponta_manually-curated_2019} is a very time
consuming process that demonstrates the need for automated tools. Both Perl et
al.~\cite{perl_vccfinder_2015} and Jimenez, Traon, and
Papadakis~\cite{jimenez_enabling_2018} tried to find fixes for specific
vulnerabilities directly in the OSS repositories in the same way: through a
reference in the NVD or through the CVE ID in the commit message.
vulnerabilities and they were able to find 1600 security patches (recall is
57\%) via these two steps. Since this research was centered around four OSS
projects, it remains unclear whether the recall will remain at this level when
looking for security fixes in different repositories.

Although we are not aware of works that focus on devising a ML model to find the
fix commit for a specific vulnerability, researchers have applied natural
language processing on the vulnerability descriptions for other applications.
Wareus and Hell~\cite{wareus_automated_2020} and Elbaz, Rilling, and
Morin~\cite{elbaz_automated_2020} both try to devise NLP and ML to automatically
label vulnerability descriptions with the Common Platform Enumeration (CPE)
labels. The former transforms the vulnerability descriptions in natural language
to a machine-readable state using the term frequency-inverse document frequency
(TF-IDF). At the same time, the latter applies word embeddings to achieve the
same goal. The CPE is an attribute provided by the NVD and is an open framework
for communicating the impact of the vulnerabilities and provides information on
the affected software version. However, there is a time lag of 35 days on
average before the vulnerability is disclosed and this attribute is assigned,
leaving automated tools unable to rely on this piece of information.

Han et al.~\cite{han_learning_2017} try to predict a vulnerability severity
score for vulnerability descriptions. In their approach, they use word
embeddings, and a one-layer shallow Convolutional Neural Network (CNN) is used
to automatically capture discriminative word and sentence features.

Similarly, Palacio et al.~\cite{palacio2019learning} proposed SecureReqNet, a
novel approach to automatically identify security-related content in issue
tracking systems using neural networks. First, high dimensional word embeddings
are extracted from vulnerability descriptions listed in the CVE database and
issue descriptions extracted from open source projects. Then, an ontology
represented by these embeddings is used to train a convolutional neural network
to predict whether a given issue is security-related.

Some related work shed light on the nature of security patches. Li and
Paxson~\cite{li2017large} conducted a large-scale empirical study concerning
security patches. Leveraging the National Vulnerability Database (NVD), they
analyze the patch development life cycle, the affected software repositories,
and security fixes to compare security fixes and non-security bug fixes. Their
results show that a third of all security issues were introduced more than three
years before their fix. Nearly 5\% of security fixes negatively impacted the
associated software, and 7\% failed to remedy the vulnerability they targeted.
Compared to \prospector, they do not use the CVE publication date but use the
earliest reference date, showing that this yields a 'significantly improved
disclosure estimation'. Given that the CVE publication is currently a
requirement of our approach, we will verify the impact of this finding on
\prospector as future work.

Finally, not all vulnerability types are the same. Therefore, Wang et
al.~\cite{wang2020machine} conducted an empirical study on vulnerability types,
using a large-scale dataset collected from the National Vulnerability Database
(NVD). Based on analysis results, they developed a classification model to
distinguish the type of security patches.

\section{Conclusion}\label{conclusion} When comparing our method to the only
similar method from Jimenez et al.~\cite{jimenez_enabling_2018} who would
achieve a precision of \precisionjimenez{}\% on the data used for this research,
which could therefore be regarded as state of the art, our implementation of our
method performed substantially better.

Through combining rather simple features that do not provide that much
information in itself into a feature vector that can be learned by ML
algorithms, we demonstrated that there lies more gain in focusing on feature
engineering instead of squeezing out a few percentages through parameter tuning
and deep learning algorithms.

The ranking results can easily be improved through adding more features or
improving current features as i.e., the lexical similarity features through the
usage of deep learning techniques and word embeddings. Furthermore,  we
presented a general method to search for fix commits in OSS repositories which
is independent of the projects and of their programming language.
\prospector, a prototype implementation of our method of mining OSS
repositories for fixes for vulnerabilities, represents a concrete first step
towards an efficient process for mining vulnerability information in open-source
code repositories. \prospector is freely available as a component of
\textsc{project ``KB''}  and is released under a liberal open-source license.

In the future, we plan to improve the filtering step by providing heuristics to
automatically find the best compromise between recall and scalability.

\begin{acks}
    This work was partly funded by project \textsc{AssureMOSS}, funded by the
    European Commission under grant number 952647.
\end{acks}

\bibliographystyle{ACM-Reference-Format}
\bibliography{references}

\newpage
\appendix
\section{Tags to tree visualization}\label{apx:tags_to_tree}

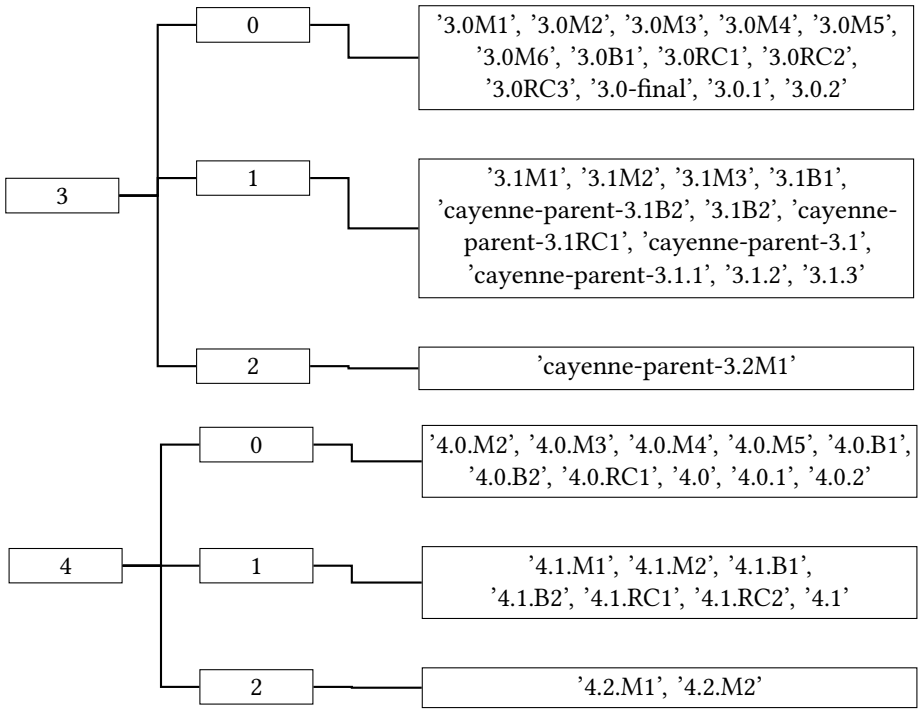
\begin{figure}[!ht]
    \textbf{Tags in the repository:}
    \begin{center}
        \texttt{['3.0-final', '3.0.1', '3.0.2', '3.0B1', '3.0M1', '3.0M2', '3.0M3', '3.0M4', '3.0M5', '3.0M6', '3.0RC1', '3.0RC2', '3.0RC3', '3.1.2', '3.1.3', '3.1B1', '3.1B2', '3.1M1', '3.1M2', '3.1M3', '4.0', '4.0.1', '4.0.2', '4.0.B1', '4.0.B2', '4.0.M2', '4.0.M3', '4.0.M4', '4.0.M5', '4.0.RC1', '4.1', '4.1.B1', '4.1.B2', '4.1.M1', '4.1.M2', '4.1.RC1', '4.1.RC2', '4.2.M1', '4.2.M2', 'cayenne-parent-3.1', 'cayenne-parent-3.1.1', 'cayenne-parent-3.1B2', 'cayenne-parent-3.1RC1', 'cayenne-parent-3.2M1']}
    \end{center}
    \mbox{}\\
    \mbox{}\\
    \textbf{Tags transformed to a tree where all tags with the same major and minor version \\ are in the same branch, and are then sorted on their timestamp:}
    \mbox{}\\
    \mbox{}\\
    \begin{tikzpicture}[grow'=right,level distance=1in,sibling distance=.25in]
        \tikzset{edge from parent/.style= {thick, draw, edge from parent fork
                right}, every level 0 node/.style= {draw,minimum
                width=0.5in,text width=0.5in,align=center}, every level 1
        node/.style= {draw,minimum width=0.5in,text
        width=0.5in,align=center}, every level 2 node/.style=
            {draw,minimum width=2.5in,text width=2.5in,align=center}, level
        2/.append style={level distance=5.5cm}} \Tree [. 3 [.{0}
        [.{'3.0M1', '3.0M2', '3.0M3', '3.0M4', '3.0M5', '3.0M6',
        '3.0B1', '3.0RC1', '3.0RC2', '3.0RC3', '3.0-final', '3.0.1',
        '3.0.2'} ]] [.1 [.{'3.1M1', '3.1M2', '3.1M3', '3.1B1',
        'cayenne-parent-3.1B2', '3.1B2', 'cayenne-parent-3.1RC1',
        'cayenne-parent-3.1', 'cayenne-parent-3.1.1', '3.1.2', '3.1.3'}
        ]] [.2 [.{'cayenne-parent-3.2M1'} ]]]
    \end{tikzpicture}
    \mbox{}\\
    \mbox{}\\
    \begin{tikzpicture}[grow'=right,level distance=1in,sibling distance=.25in]
        \tikzset{edge from parent/.style= {thick, draw, edge from parent fork
                right}, every level 0 node/.style= {draw,minimum
                width=0.5in,text width=0.5in,align=center}, every level 1
        node/.style= {draw,minimum width=0.5in,text
        width=0.5in,align=center}, every level 2 node/.style=
            {draw,minimum width=2.5in,text width=2.5in,align=center}, level
        2/.append style={level distance=5.5cm}} \Tree [. 4 [.{0}
        [.{'4.0.M2', '4.0.M3', '4.0.M4', '4.0.M5', '4.0.B1', '4.0.B2',
        '4.0.RC1', '4.0', '4.0.1', '4.0.2'} ]] [.1 [.{'4.1.M1',
        '4.1.M2', '4.1.B1', '4.1.B2', '4.1.RC1', '4.1.RC2', '4.1'} ]]
        [.2 [.{'4.2.M1', '4.2.M2'} ]]]
    \end{tikzpicture}

    \caption{The result of transforming tags for the GitHub repository Apache Cayenne to a sorted tags tree, whereby the inconsistent versioning is ignored (e.g., the usage of 'cayenne-parent' and tags with only major.minor).}
    \label{fig:tags_to_tags_tree}
\end{figure}

\end{document}